\documentclass[a4paper,10pt]{spieman}  % use this instead for A4 paper
\usepackage{amsmath,amsfonts,amssymb}
\usepackage{graphicx}
\usepackage{setspace}
\usepackage{tocloft}
\usepackage[table,xcdraw]{xcolor}
\usepackage{tabularx,colortbl}
\usepackage{booktabs}
\usepackage{amsmath}
\usepackage{color,soul}
\DeclareMathOperator{\atantwo}{atan2}

\usepackage{threeparttable}
\usepackage{longtable}
\usepackage{geometry}
\usepackage{booktabs}
\usepackage{lscape}
\usepackage{array}

\makeatletter
\renewcommand{\maketitle}{%
  \null
  \vspace{0.2in}
  \begin{center}
    {\large\bfseries \@title \par}
    \vskip 1.0em
    \@author
    \vskip 1.0em
    \@date
  \end{center}
}
\makeatother

\title{Assessment of the AlUla Manara astronomical site in Saudi Arabia using ECMWF ERA5 Reanalysis data}

\author[a,b,c, *]{Gianluca Lombardi}
\author[b,**]{Gary Fildes}
\author[d]{Omar Cuevas}
\author[b]{Thamer Y. Alrefay}
\author[b]{Naif Almalik}
\affil[a]{GRANTECAN S.A., Cuesta de San Jos\'e s/n, E-38712 Bre\~na Baja, La Palma, Spain}
\affil[b]{Royal Commission for AlUla, Al Safarat 12512, Riyadh, Saudi Arabia}
\affil[c]{Instituto de Astrof\'isica de Canarias, E-38205 La Laguna, Tenerife, Spain}
\affil[d]{Instituto de F\'isica y Astronom\'ia, Universidad de Valpara\'iso, Gran Breta\~na 1111, Valpara\'iso, Chile}

\cftpagenumbersoff{figure}
\cftpagenumbersoff{table} 
\begin{document}

\noindent\textit{Preliminary version. Paper submitted to JATIS (Journal of Astronomical Telescopes, Instruments, and Systems) and accepted for publication on 18 July 2025.}

\vspace{0.5cm}

\noindent\textit{The AlUla Manara Observatory project is managed by the \textbf{Royal Commission for AlUla} as part of Saudi Arabia Vision 2030 and will be the first cutting-edge observational facility in the Gulf Countries region.}

\vspace{-1em}

\maketitle

\begin{abstract}
As part of Saudi Arabia Vision 2030 and under the guidance of the Royal Commission for AlUla (RCU), efforts are underway to establish AlUla Manara as the Kingdom first major astronomical observatory. This study presents a preliminary assessment of the site based on ECMWF ERA5 reanalysis data to evaluate its suitability for hosting a 4m-class optical-IR telescope. AlUla Manara is located on a remote plateau 74 km north of the historical town of AlUla and was recently designated as an International Dark Sky Park. The analysis focuses on key astro-meteorological parameters such as seeing, temperature regimes, wind patterns, cloud cover and precipitable water vapor (PWV). Results show a median nighttime seeing of 1.5 arcsec, a median cold season PWV of 3.2 mm, and over 79\% of nighttime hours with clear sky conditions. Wind regimes are generally mild, posing no constraints on infrastructure. The analysis includes three further sites in the Kingdom, namely Volcanic Top, Ward Mountain, and Dubba Mountain. These sites exhibit better turbulence conditions, but are located outside RCU jurisdiction. Nevertheless, AlUla Manara remains a competitive candidate thanks to its alignment with broader regional development goals. To validate these preliminary results, a dedicated Astronomical Site Monitor has been deployed on site to support the design and operational planning of the observatory.
\end{abstract}

% Include a list of up to six keywords after the abstract

% Include email contact information for corresponding author
{\noindent \footnotesize\textbf{*} Gianluca Lombardi, \linkable{gianluca.lombardi@gtc.iac.es} }\\
{\noindent \footnotesize\textbf{**} Gary Fildes, \linkable{g.fildes@rcu.gov.sa} }

\begin{spacing}{1}   % use double spacing for rest of manuscript

\section{Introduction}
\label{sect:intro}  % \label{} allows reference to this section
AlUla is an historic region in the province of Medinah, in the North-West of the Kingdom of Saudi Arabia, renowned for its extensive cultural heritage, including ancient archaeological sites, unique rock formations, and well-preserved Nabatean tombs and inscriptions. The old Town of AlUla, with its traditional mud-brick buildings and narrow alleyways, is a notable feature of this heritage-rich area. In recent years, AlUla has drawn attention for its ambitious development projects, which seek to preserve its heritage while promoting tourism and economic growth in alignment with Kingdom’s Vision 2030 goals. The Royal Commission for AlUla (RCU) is focused on a sustainable transformation of the region, prioritizing environmental and historical conservation.\\
A key initiative under this vision is the development of a distinguished astronomical observatory, combining astro-tourism with advanced astronomical research, equipping the site with a 4m-class optical-IR telescope. Recently, Ref. \citenum{Alrefay2024} analyzed the geographical and geological characteristics of several locations within the RCU jurisdiction. In particular, the study focused on ground stability and site accessibility, ultimately identifying AlUla Manara as the location for Saudi Arabia’s first astronomical observatory.\\
AlUla Manara (\textit{The Lighthouse} in Arabian) is a wide plateau in a remote rocky desert located about 74 km North of AlUla. As a demonstration of the commitment of the RCU with the astronomical development of the region, in October 2024, the AlUla Manara and Al Gharameel Nature Reserves have been accredited as International Dark Sky Parks by the DarkSky International authority\footnote{\linkable{https://darksky.org/places/alula-manara-and-algharameel-nature-reserves/}}.\\
To fully assess the properties of AlUla Manara, an Astronomical Site Monitor (ASM) has been deployed at the site during the first trimester of 2025 for a comprehensive characterization. The ASM started operations at the end of April 2025, and will initially operate for a period of one year, to cover eventual differences between seasons.\\
Preliminary analysis presented in (\citenum{lombardi2024}), used a limited subset of remote sensing data from the European Centre for Medium-Range Weather Forecasts (ECMWF) Fifth Generation Reanalysis (ERA5) covering the period 2018--2023 (\citenum{Hersbach2020}). That initial study relied on parameters with temporal resolutions ranging from 1 to 6 hours. In the present work, we expand and complete the assessment by incorporating ERA5 data spanning a significantly longer time-frame, from 1990 to 2024 (35 years), and by uniformly enhancing the temporal resolution to 1 hour. Furthermore, a refined data processing strategy has been implemented to ensure a robust study and enabling a more comprehensive characterization of the astro-meteorological conditions at the site. The results presented hereafter show some differences compared to those reported in (\citenum{lombardi2024}), although the general seasonal trends are confirmed.\\
\begin{figure}
\begin{center}
\begin{tabular}{c}
\includegraphics[height=7.5cm]{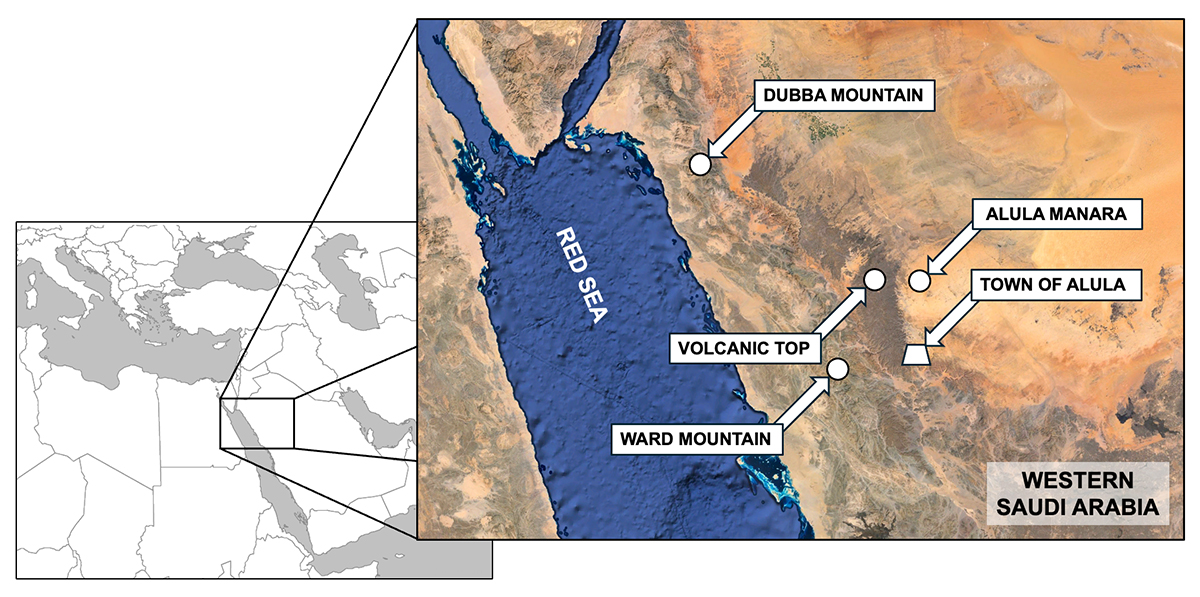}
\end{tabular}
\end{center}
\caption 
{\label{fig:sites}
Map of the location of the four sites evaluated in this study, together with the AlUla Old Town.}
\end{figure}
\begin{table}
\caption{Geographical locations of the sites.} 
\label{tab:sites}
\begin{center}
\small
\begin{tabular}{|r|c|c|c|c|} 
\hline
\rule[-1ex]{0pt}{3.5ex}   & \textbf{Latitude} & \textbf{Longitude} & \textbf{Elevation} & \textbf{distance from} \\
\rule[-1ex]{0pt}{3.5ex}   &  &  &  & \textbf{AlUla Manara} \\
\hline\hline
\rule[-1ex]{0pt}{3.5ex}  \textbf{AlUla Manara} & 27\textdegree11'32.4"N & 37\textdegree48'40.1"E & 1209 m.a.s.l. & --\\
\hline
\rule[-1ex]{0pt}{3.5ex}  \textbf{Volcanic Top} & 27\textdegree09'42.0"N  & 37\textdegree29'08.3"E  & 1807 m.a.s.l. & $\sim$33 km (W)\\
\hline
\rule[-1ex]{0pt}{3.5ex}  \textbf{Ward Mountain} & 26\textdegree24'47.7"N  & 37\textdegree13'08.8"E & 2052 m.a.s.l. & $\sim$104 km (SW)\\
\hline
\rule[-1ex]{0pt}{3.5ex}  \textbf{Dubba Mountain} & 27\textdegree51'45.7"N  & 35\textdegree44'23.4"E & 2146 m.a.s.l. & $\sim$218 km (NW)\\
\hline
\end{tabular}
\end{center}
\end{table}
For comparative analysis, three additional locations within Saudi Arabia -- Volcanic Top, Ward Mountain, and Dubba Mountain -- have been included in the study due to their apparent favorable geographical positions for astronomical observatories. These sites are located between $\sim$33 km and $\sim$218 km from AlUla Manara and lie outside the RCU jurisdiction. Their inclusion has a dual purpose: first, they provide a contextual benchmark to assess the performance of the AlUla Manara plateau within a national framework; second, although no in-situ instrumentation is currently planned for any of the additional sites, this preliminary scientific assessment may highlight candidate sites for future large-aperture projects in the region.\\
The general properties of the sites are reported below, while geographical details are provided in Table \ref{tab:sites} and Figure \ref{fig:sites}:
\begin{itemize}
\item \textbf{AlUla Manara}. The main astronomical site, inside the RCU jurisdiction area. It is part of a large wide plateau with an extension of approximately 60 km $\times$ 100 km, with an elevation of about 1200 m.a.s.l. Its Nature Reserve has been accredited as International Dark Sky Park.
\item \textbf{Volcanic Top}. Outside the RCU jurisdiction area. It's the highest point (1807 m.a.s.l.) of a large volcanic plateau (approx. 30 km $\times$ 100 km) in North-West to South-East direction at $\sim$33 km in western direction from AlUla Manara. The area is rich in archaeological and geological finds, thus it could not be available for astronomical research development.
\item \textbf{Ward Mountain}. Outside the RCU jurisdiction area. It is a peak on the western mountain ridge of Saudi Arabia, in proximity of the Red Sea shore, at $\sim$104 km in southwestern direction from AlUla Manara.. It has an elevation of 2052 m.a.s.l. and currently hosts some sort of infrastructure. Its availability for astronomical scientific research is unclear.
\item \textbf{Dubba Mountain}. Outside the RCU jurisdiction area. Dubba is a steep mountain with an elevation of 2146 m.a.s.l. in the North-West of Saudi Arabia at $\sim$218 km in northwestern direction from AlUla Manara. Its accessibility is not yet assessed.
\end{itemize}
The results from the ongoing site characterization efforts will play a critical role in the design, construction, and operation of the future 4m-class telescope. The results will also help to establish a reasonable scientific direction for the future observatory. As of now, there is no available local data for the parameters to be characterized at any of the sites, with the exception of AlUla Manara, where the ASM operations have just started. This study aims to provide an initial assessment of previously unexplored sites, offering the astronomical scientific team at RCU the necessary insights to guide further evaluation and ensure that AlUla Manara is effectively utilized for both national and international astronomical research.
A comprehensive, long-term analysis using local data collected by the instruments deployed within the ASM at AlUla Manara will ultimately validate or challenge these initial conclusions. Until then, this work serves as a provisional foundation for assumptions regarding the designated astronomical site.\\
The paper is structured as follows:
\begin{itemize}
\item[$-$] \textbf{Section \ref{sect:data}}: describes the datasets and analysis methodologies.
\item[$-$] \textbf{Section \ref{sect:seeing}}: describes the characterization of the atmospheric turbulence.
\item[$-$] \textbf{Section \ref{sect:t}}: describes the ground temperature properties.
\item[$-$] \textbf{Section \ref{sect:wind}}: describes the wind speed and wind direction patterns.
\item[$-$]\textbf{Section \ref{sect:tcc}}: describes the estimation of cloud coverage and fraction of hours with clear sky.
\item[$-$] \textbf{Section \ref{sect:pwv}}: describes the Precipitable Water Vapor (PWV) prediction.
\item[$-$] \textbf{Section \ref{sect:asm}}: disclose a brief description of the deployment of the Astronomical Site Monitor at AlUla Manara.
\item[$-$] \textbf{Section \ref{sect:conclusions}}: reports the Conclusions of this study.
\end{itemize}

\begin{table}
\begin{center}
\small
\begin{threeparttable}
\caption{Downloaded surface-level and vertical layers metrics.}
\label{tab:metrics}
\renewcommand{\arraystretch}{1.4}
\begin{tabular}{|>{\raggedleft\arraybackslash}p{0.16\textwidth}%
                >{\raggedright\arraybackslash}p{0.32\textwidth}|
                >{\raggedleft\arraybackslash}p{0.16\textwidth}%
                >{\raggedright\arraybackslash}p{0.32\textwidth}|}
\hline
\multicolumn{2}{|c|}{\textbf{SURFACE-LEVEL}} & \multicolumn{2}{c|}{\textbf{VERTICAL LAYERS}} \\
\noalign{\vskip -4pt}
\multicolumn{2}{|c|}{\textbf{ERA5-Land}} & \multicolumn{2}{c|}{\textbf{ERA5}} \\
\hline\hline
\textit{Temporal res.:} & \hspace{-3pt}1 hour & \textit{Temporal res.:} & \hspace{-3pt}1 hour \\
\textit{Spatial res.:} & \hspace{-3pt}0.10$^\circ \times$ 0.10$^\circ$  & \textit{Spatial res.:} & \hspace{-3pt}0.25$^\circ \times$ 0.25$^\circ$ \\
\noalign{\vskip -6pt}
 & \hspace{-3pt}9 km $\times$ 9 km (approx.) &  & \hspace{-3pt}27.8 km $\times$ 27.8 km (approx.)\\
 &  & \textit{Vertical res.:} & \hspace{-3pt}10 -- 1000 hPa \\
\noalign{\vskip -6pt}
 & \hspace{-3pt} &  & \hspace{-3pt}{\scriptsize [10, 20, 30, 50, 70, 100, 125, 150, 175, 200, 225, 250, 300, 350, 400, 450, 500, 550, 600, 650, 700, 750, 775, 800, 825, 850, 875, 900, 925, 950, 975, 1000]} \\
\hline\hline
\multicolumn{2}{|l|}{Boundary layer height} & \multicolumn{2}{l|}{Geopotential} \\
\hline
\multicolumn{2}{|l|}{2 m temperature} & \multicolumn{2}{l|}{Temperature} \\
\hline
\multicolumn{2}{|l|}{10 m (\textit{u}, \textit{v}) wind components} & \multicolumn{2}{l|}{(\textit{u}, \textit{v}) wind components} \\
\hline
\multicolumn{2}{|l|}{Total cloud cover} & \multicolumn{2}{l|}{Total column water vapor\tnote{(*)}} \\
\hline
\end{tabular}
\begin{tablenotes}
\footnotesize
\item[(*)] The ERA5 \textit{total column water vapour} variable is not a surface-level or vertically resolved variable, it's a single-level, vertically integrated variable.
\end{tablenotes}
\end{threeparttable}
\end{center}
\end{table}

\section{Data and Methodology}
\label{sect:data}
\subsection{Datasets}\label{sect:datasets}
The results presented in this study are based on the analysis of reanalysis datasets produced by the ECMWF. Specifically, surface-level variables were obtained from the ERA5--Land dataset, while vertical profile data were sourced from the ERA5 reanalysis (\citenum{Hersbach2020}; \citenum{MunozSabater2021}). Both datasets are retrievable with a temporal resolution of 1 hour.\\
Although ECMWF is headquartered in Europe, its reanalysis products integrate a wide range of global observations, including contributions from NASA, JAXA, and other international agencies.\\
The ERA5 atmospheric reanalysis delivers a horizontal resolution of 0.25$^\circ \times$ 0.25$^\circ$ (about 27.8 km $\times$ 27.8 km at the equator), and supplies vertical profiles on 137 model levels extending from the surface up to 0.01 hPa. This grid spacing refers to the intervals at which data are output and made available to users.\\
ERA5--Land offers enhanced spatial detail for land surface variables, providing data at a horizontal resolution of 0.10$^\circ \times$ 0.10$^\circ$, which corresponds to approximately 9 km $\times$ 9 km at the equator (\citenum{MunozSabater2021}). ERA5--Land variables are referenced as above the local land surface elevation, rather than above sea level (\citenum{MunozSabater2021}). Surface values provided by ERA5–Land are calculated through vertical interpolation from model levels to standard reference heights (e.g., 2 m for temperature, 10 m for wind), based on near-surface atmospheric profiles and surface layer parameterizations (\citenum{Hersbach2020}; \citenum{MunozSabater2021}).\\
The specific surface-level and vertical-layer variables retrieved for this study are summarized in Table \ref{tab:metrics}, together with details about the temporal, spatial and vertical resolutions of the data. Surface-level metrics were exclusively taken from ERA5--Land to leverage its finer spatial resolution over land areas. Vertical atmospheric parameters were extracted from the ERA5 dataset.\\
For all retrieved data the temporal resolution is of 1 hour, and the covered period time-frame is between 1 January 1990 to 31 December 2024 (35 years), with the exception of the seeing estimation for which the covered period extends only for nighttime data between 1 January 2015 to 31 December 2024 (10 years). A thorough check confirmed the absence of temporal gaps for all downloaded datasets.\\
Something worth considering are certain limitations when applying ERA5 and ERA5-Land datasets to astronomical site assessment. These reanalysis products are typically used in climatology and atmospheric studies, but their spatial resolution (27.8 km for ERA5, 9 km for ERA5-Land) may not fully capture local terrain effects that influence turbulence and cloud development. The vertical resolution, while generally sufficient for mesoscale analyses, can under-represent turbulence near the surface, especially within the Planetary Boundary Layer (PBL), an aspect particularly relevant for atmospheric turbulence estimations. Furthermore, since the datasets are deterministic, they do not inherently provide formal uncertainty ranges. Despite these limitations, ERA5 offers a valuable first approximation of the site's long-term atmospheric behavior.
\begin{table}
\caption{Definitions of \textit{daytime} and \textit{nighttime}, as well as of the Saudi Arabia \textit{warm season} and \textit{cold season} months ranges.} 
\label{tab:def}
\begin{center}
\small
\begin{tabular}{|r|l|} 
\hline
\rule[-1ex]{0pt}{3.5ex} \textbf{Daytime}  & \textit{sunrise} to \textit{sunset} \\
\hline
\rule[-1ex]{0pt}{3.5ex} \textbf{Nighttime}  & \textit{sunset} to \textit{sunrise} \\
\hline
\rule[-1ex]{0pt}{3.5ex} \textbf{Warm Season}  & April to September \\
\hline
\rule[-1ex]{0pt}{3.5ex} \textbf{Cold Season}  & October to March next year \\
\hline
\end{tabular}
\end{center}
\end{table} 
\subsection{Data analysis methodology}\label{sect:method}
Table \ref{tab:def} reports the definitions of \textit{daytime} and \textit{nighttime}, as well as months ranges for the Saudi Arabia \textit{warm season} and \textit{cold season}. These definitions have been widely used during the analysis.
Surface-level variables such as \textit{2m temperature}, \textit{10 m wind components}, \textit{total cloud cover}, plus the vertical integrated \textit{total column water vapor}, will be treated similarly, with subtle differences that may depend on specific properties we want to analyze. These specific cases are mentioned in their respective sections in the Paper. For what concerns the atmospheric turbulence, in Section \ref{sect:seeing} we describe in details the model implemented to calculate the seeing for the four locations under study. Once the seeing is retrieved, it will be treated with the same methodology as any other downloaded surface-level variable.\\
Having all variables available with 1 hour resolution, we are able to properly separate data in daytime from data in nighttime. This is achieved with a dedicated Python script that calculates the precise sunset and sunrise times based on the sites' coordinates, timezone and elevation above sea level. As an additional information, evening and morning twilight times are included in the nighttime. The distinction allows to better distinguish with improved precision eventual differences in the properties of the sites between daytime and nighttime.\\
For every variable $x$ we may select data in connection with daytime, nighttime and/or among 24-hours. In every case we have calculated the median and/or the average $x$ for every month $m$ in a specific year $y$, calling them $med(x)_{m,y}$ and $avg(x)_{m,y}$ respectively. Afterwards, the medians and/or the averages in the same month $m$ in different years $y$ are averaged with the formulas
\begin{equation}
\overline{med(x)_{m}} = \frac{1}{N} \sum_{y=start}^{2024} med(x)_{m,y},
\label{eq:monthly_med}
\end{equation}
\begin{equation}
\overline{avg(x)_{m}} = \frac{1}{N} \sum_{y=start}^{2024} avg(x)_{m,y}
\label{eq:monthly_avg}
\end{equation}
For all variables the starting year is 1990, and $N = 35$ is total number of years covered by the datasets, with the exception of the seeing where the starting year is 2015 and $N = 10$.\\
With equations \ref{eq:monthly_med} and \ref{eq:monthly_avg} we calculate what we define as the \textit{time-frame monthly averages}, where $\overline{med(x)_{m}}$ represents literally the \textit{average of the medians of a specific month through the whole time-frame}, while $\overline{avg(x)_{m}}$ is \textit{the average of the monthly-averages of specific month through the whole time-frame}. In both cases, the respective standard deviations $\sigma_{med,y}$ and $\sigma_{avg,y}$ are also calculated. The averages $\overline{med(x)_{m}} \pm \sigma_{med,y}$ and $\overline{avg(x)_{m}}  \pm \sigma_{avg,y}$ will allow to assess the general seasonal trend of a variables through the entire time-frame.\\
In addition to the standard deviations, for the analysis we have added a further parameter called \textit{fluctuation}. We define the \textit{fluctuation} of a variable $x$ as the average range of the differences between its maximum and minimum values observed within a specific \textit{time interval} and aggregated over longer periods. The procedure to calculate the fluctuation is described as follows:
\begin{itemize}
\item[$-$]\textbf{\textit{Short} time interval fluctuation}\\
This refers to the intra-day variability of a given variable. For a specific month $m$, the range is computed as the difference between the maximum and minimum values recorded within a defined short time interval (daytime, nighttime, or the full 24-hour cycle). These daily fluctuations are then averaged over the month. These monthly ranges are then averaged over the years (i.e., all Januaries, all Februaries etc.):
\begin{equation}
\text{FL}_{m}^{\text{short}}(x) = \frac{1}{N} \sum_{y=1}^{N} \left[ \frac{1}{D_{m,y}} \sum_{i=1}^{D_{m,y}} \left( x_{\max,i}^{(m,y)} - x_{\min,i}^{(m,y)} \right) \right]
\label{eq:short_fluctuation}
\end{equation}
where $x_{\max,i}^{(m,y)}$ and $x_{\min,i}^{(m,y)}$ are the maximum and minimum values of variable $x$ for the $i$-th day (or night) of month $m$ in year $y$, $D_{m,y}$ is the number of intervals in that month, and $N$ is the total number of years considered.
\vspace{0.5em}
\item[$-$]\textbf{\textit{Long} time interval fluctuation}\\
In this case, the fluctuation is computed across an entire calendar month for each year. For each year, the maximum and minimum values of the variable within the month are determined, and the difference is taken. These monthly ranges are then averaged over the years (i.e., all Januaries, all Februaries etc.):
\begin{equation}
\text{FL}_{m}^{\text{long}}(x) = \frac{1}{N} \sum_{y=1}^{N} \left( x_{\max, m}^{(y)} - x_{\min, m}^{(y)} \right)
\label{eq:long_fluctuation}
\end{equation}
where \( x_{\max, m}^{(y)} \) and \( x_{\min, m}^{(y)} \) are the maximum and minimum values of the variable \( x \) within calendar month \( m \) of year \( y \), and \( N \) is the total number of years considered.
\end{itemize}
This fluctuation metric allows us to capture the typical intra-period variability of the variable and is used uniformly across different atmospheric and astro-meteorological quantities, such as the seeing, day/night temperature gradients, and cloud coverage, providing a consistent way to assess their dynamic range within natural cycles.\\
For what concerns specific cases, we have considered the \textit{short} time interval fluctuation appropriate for the seeing (i.e. difference between maximum and minimum during a single night), while \textit{long} time interval fluctuation is applied to cases like the day/night temperature gradients and the cloud coverage (maximum and minimum during a single month).\\
All the concepts and equations described in this section will be widely used in the next sections, enabling a consistent and integrated assessment of the astro-meteorological conditions across the sites.
\section{Atmospheric Turbulence}\label{sect:seeing}
\subsection{Estimation of $C_{n}^{2}$}
The $C_{n}^{2}$ is the refractive index structure coefficient that represents the distribution of the turbulent energy from the ground to the stratosphere, and it varies with temperature, pressure and humidity fluctuations, causing light to refract as it propagates through non-homogeneous mediums (\citenum{Cuevas2024}). The $C_{n}^{2}$ is used to characterize the quality of images collected by astronomical observatories.\\
Several methods can be found in the literature to estimate the astro-meteorological parameters based on atmospheric variables, some of them used ERA5 data to characterize astronomical sites (\citenum{Bi2023}; \citenum{Bekhard2024}; \citenum{Shikhovtsev2024}). Authors in (\citenum{Osborn2018}) used a model based on Gladstone's equation to estimate the $C_{n}^{2}$ vertical profile and seeing over Cerro Paranal site in Chile. We used a similar methodology as a standard approximation that can be used to characterize new astronomical sites, where observations are not available. The Gladstone equation estimates the $C_{n}^{2}$ profile based on atmospheric variables such as temperature [K], pressure [hPa] and the temperature structure constant ($C_{T}^{2}$). The authors in (\citenum{Masciadri2017}) advise to use a modification in Gladstone's equation because it assumes that the atmosphere is in hydrostatic equilibrium and the temperature gradient follows the adiabatic approximation, but in some parts of the atmosphere (recurrently at high levels), the difference cannot be negligible. Tatarskii (\citenum{tatarskii1971}) demonstrated that a more general and formally correct replacement for temperature $T$ is the potential temperature $\theta$, which is a conservative quantity under adiabatic conditions. This modified version of the Gladstone equation (Gm afterwards) was used by (\citenum{Coulman:88}; \citenum{Abahamid2004}; \citenum{Cherubini2013}; \citenum{Masciadri2017}; \citenum{Osborn2018}; \citenum{Cuevas2024}). Therefore, $T^2$ is replaced by $T\theta$ using the following expression:
\begin{equation}
\label{eq:G_1}
C_{n}^{2}=\left( \frac{80\times10^{-6}P}{T{\theta}}\right)^{2} C_{T}^{2} \, ,
\end{equation}
where $\theta$ is the potential temperature in [K], which is estimated by
\begin{equation} \label{eq:G_2}
\theta = T\left( \frac{P_{0}}{P}\right)^{\frac{R}{c_{p}}} \, ,
\end{equation}
where $P_{0}$ is a standard reference pressure, usually 1000 hPa, $P$ is the pressure in the vertical, $R$ is the gas constant of air, and $c_p$ is the specific heat capacity at a constant pressure. In meteorology, $R/c_{p}$ = 0.286. The temperature structure coefficient ($C_{T}^{2}$) along the vertical path $L$ is defined as:
\begin{equation}
\label{eq:G_3}
C_{T}^{2}=L^{\frac{4}{3}}\left(\frac{\partial{\theta}}{\partial{z}}\right)^{2} \phi \, ,
\end{equation}
The $\phi$ parameter represents the thermal and dynamic stability of the atmosphere (\citenum{Masciadri2001}) and $L$ can be expressed as a function of $\theta$ and the turbulent kinetic energy ($TKE$) using:
\begin{equation}
\label{eq:G_4}
L=\sqrt{\frac{2TKE}{\frac{g\partial{\theta}}{\theta\partial{z}}}} \, ,
\end{equation}
where $g$ is the acceleration due to gravity. Similar to (\citenum{Osborn2018}), the $TKE$ was calculated as $TKE=S^{2}$, where $S$ is the vertical wind shear calculated by:
\begin{equation}
\label{eq:G_5}
    S=\left[\left( \frac{\partial{u}}{\partial{z}}\right)^2 +\left( \frac{\partial{v}}{\partial{z}}\right)^2 \right] ^\frac{1}{2}
\end{equation}
and $u$ and $v$ are the x- and y- (zonal and meridional) wind components, respectively, obtained from ERA5 database.\\
Finally, replacing equation \ref{eq:G_4} in \ref{eq:G_3}, the final expression for $C_{n}^{2}$ is:
\begin{equation}
    \label{eq:G_6}
    C_{n}^{2}=\phi \left(\frac{80\times10^{-6}P}{T\theta}\right)^{2}L^{\frac{4}{3}}\left(\frac{\partial{\theta}}{\partial{z}}\right)^{2}
\end{equation}
The equation represents the Gladstone modified (Gm) model. The function $\phi$ is used to calibrate the $C_{n}^{2}$ profile when observational data is available (\citenum{Osborn2018}; \citenum{Cuevas2024}). Additionally, (\citenum{Cuevas2024}) demonstrates that the modified Gladstone model (\ref{eq:G_6}) underestimates the $C_{n}^{2}$ profile at Paranal. Similarly to (\citenum{Osborn2018}), we applied a constant value for $\phi$ to uniformly increase the $C_{n}^{2}$ values across the entire profile. We have chosen $\phi = 6$ because, when using the same methodology, this value has provided the best match between our simulated $C_{n}^{2}$ profiles and G-SCIDAR (Generalized SCIDAR) observations taken at the Roque de Los Muchachos Observatory (ORM) in La Palma (\citenum{GSCIDAR}). Although our observational results at ORM are not yet published, this value consistently led to an increase in $C_{n}^{2}$ that best reflects the measured data. The sites we analyze in this work lie at a similar latitude to ORM, and while their local climate may differ, we expect that the large-scale atmospheric circulation in the free atmosphere is likely to be comparable.
\subsection{Seeing estimation}
Astronomical seeing is a parameter that quantifies the blurring and twinkling of celestial objects caused by atmospheric turbulence, and it serves as an indicator of the quality of astronomical sites. One common method to estimate astronomical seeing is by calculating the Fried parameter , $r_{0}$, which involves vertically integrating the refractive index structure parameter, $C_{n}^{2}$ (\ref{eq:G_6}), using the expression provided in (\citenum{Roddier1981}; \citenum{Roddier1989}):
\begin{equation}
\label{eq:r0}
r_{0}=\left[0.423\left(\frac{2\pi}{\lambda}\right)^2\int_{0}^{L}C_{n}^{2}(z)d(z)\right]^{-\frac{3}{5}} \, ,
\end{equation}
where $\lambda$ is the wavelength at which the telescope is observing and $L$ is the total path length. Estimation of seeing ($\epsilon$) is obtained using the relation:
\begin{equation}
\label{eq:seeing}
\epsilon=0.98\frac{\lambda}{r_{0}}
\end{equation}
From the formulas we understand that the seeing is inversely proportional to the wavelength of observation. At longer wavelength it will affect images less than at shorter wavelengths.
\begin{table}
\newcolumntype{P}[1]{>{\centering\arraybackslash}p{#1}}
\caption{Seeing statistics in [arcsec] from the Gm model in the visible ($\lambda$ = 550 nm).}
\label{tab:seeing}
\renewcommand{\arraystretch}{1.2}
\begin{center}       
\small
\begin{tabular}{|>{\columncolor{gray!15}}c|r|P{1.2cm}|P{1.2cm}|P{1.2cm}|c|P{1.2cm}|P{1.2cm}|} 
\hline
  \cellcolor{white} & & \textbf{min} & \textbf{25\%} & \textbf{median} & \textbf{average} & \textbf{75\%} & \textbf{max}\\  
\hline
\noalign{\vskip 3pt}
\hline
  \textbf{AlUla}    & Annual      & 0.3 & 1.1 & 1.5 & 1.6 $\pm$ 0.7 & 2.0 & 5.7 \\  
\cline{2-8}
  \textbf{Manara}   & Warm Season & 0.3 & 0.9 & 1.3 & 1.4 $\pm$ 0.6 & 1.7 & 4.1 \\  
                    & Cold Season & 0.3 & 1.2 & 1.7 & 1.8 $\pm$ 0.7 & 2.3 & 5.7 \\
\hline
\noalign{\vskip 3pt}
\hline
  \textbf{Volcanic} & Annual      & 0.3 & 0.8 & 1.1 & 1.3 $\pm$ 0.6 & 1.6 & 5.8 \\  
\cline{2-8}
  \textbf{Top}      & Warm Season & 0.3 & 0.7 & 0.9 & 1.0 $\pm$ 0.5 & 1.2 & 4.2 \\  
                    & Cold Season & 0.3 & 1.0 & 1.4 & 1.5 $\pm$ 0.7 & 1.8 & 5.8 \\
\hline
\noalign{\vskip 3pt}
\hline
  \textbf{Ward}     & Annual      & 0.2 & 0.7 & 0.9 & 1.0 $\pm$ 0.4 & 1.2 & 3.4 \\  
\cline{2-8}
  \textbf{Mountain} & Warm Season & 0.2 & 0.6 & 0.7 & 0.8 $\pm$ 0.3 & 0.9 & 3.0 \\  
                    & Cold Season & 0.3 & 0.9 & 1.1 & 1.1 $\pm$ 0.4 & 1.4 & 3.4 \\
\hline
\noalign{\vskip 3pt}
\hline
  \textbf{Dubba}    & Annual      & 0.2 & 0.7 & 0.8 & 0.9 $\pm$ 0.4 & 1.1 & 3.7 \\  
\cline{2-8}
  \textbf{Mountain} & Warm Season & 0.2 & 0.6 & 0.7 & 0.8 $\pm$ 0.2 & 0.9 & 2.6 \\  
                    & Cold Season & 0.3 & 0.8 & 1.1 & 1.1 $\pm$ 0.4 & 1.3 & 3.7 \\  
\hline
\end{tabular}
\end{center}
\end{table}
\subsection{Model outcome}\label{sect:outcome}
In the present study the seeing estimations have been processed for nighttime data (\textit{sunset} to \textit{sunrise} (see Table \ref{tab:def}). The seeing has been calculated in the visible ($\lambda$ = 550 nm) with the use of equation \ref{eq:G_6}  (Gm model) on the path $L$ considering the vertical levels reported in Table \ref{tab:metrics} with a time resolution of 1 hour, allowing for full statistics between 2015 and 2024 (see Table \ref{tab:seeing}). In the table, seeing values are reported with a resolution of 0.1 arcsec, as increasing the apparent numerical precision to 0.01 arcsec (i.e., 1.47 vs. 1.5 arcsec) would not be meaningful at this stage. The accuracy inherent in the model-based estimation of  $r_{0}$ combined with the exploratory nature of this study --which aims to identify general trends rather than provide deterministic forecasts-- do not warrant such a level of detail. At this early stage, in the absence of direct observational references, we are not in a position to quantify the model’s uncertainty. However, we acknowledge that its precision is limited. Based on findings from similar studies (\citenum{Masciadri2017}; \citenum{Osborn2018}), we expect the model to more accurately represent conditions in the upper atmosphere, while being less reliable in the lower layers. \\
From the analysis, AlUla Manara presents a median of 1.5 arcsec, while Ward Mountain and Dubba Mountain with their elevations higher of about 800 m above AlUla Manara are characterized by more favorable turbulence conditions, with medians of 0.9 and 0.8 arcsec respectively, in line with world-class astronomical observatories such as Cerro Paranal (0.72 arcsec; \citenum{par2024}) and the ORM (0.80 arcsec; \citenum{orm2012}). Finally, the Volcanic Top has middling conditions, with a median of 1.1 arcsec.\\
\begin{figure}
\begin{center}
\begin{tabular}{c}
\includegraphics[width=0.98\textwidth]{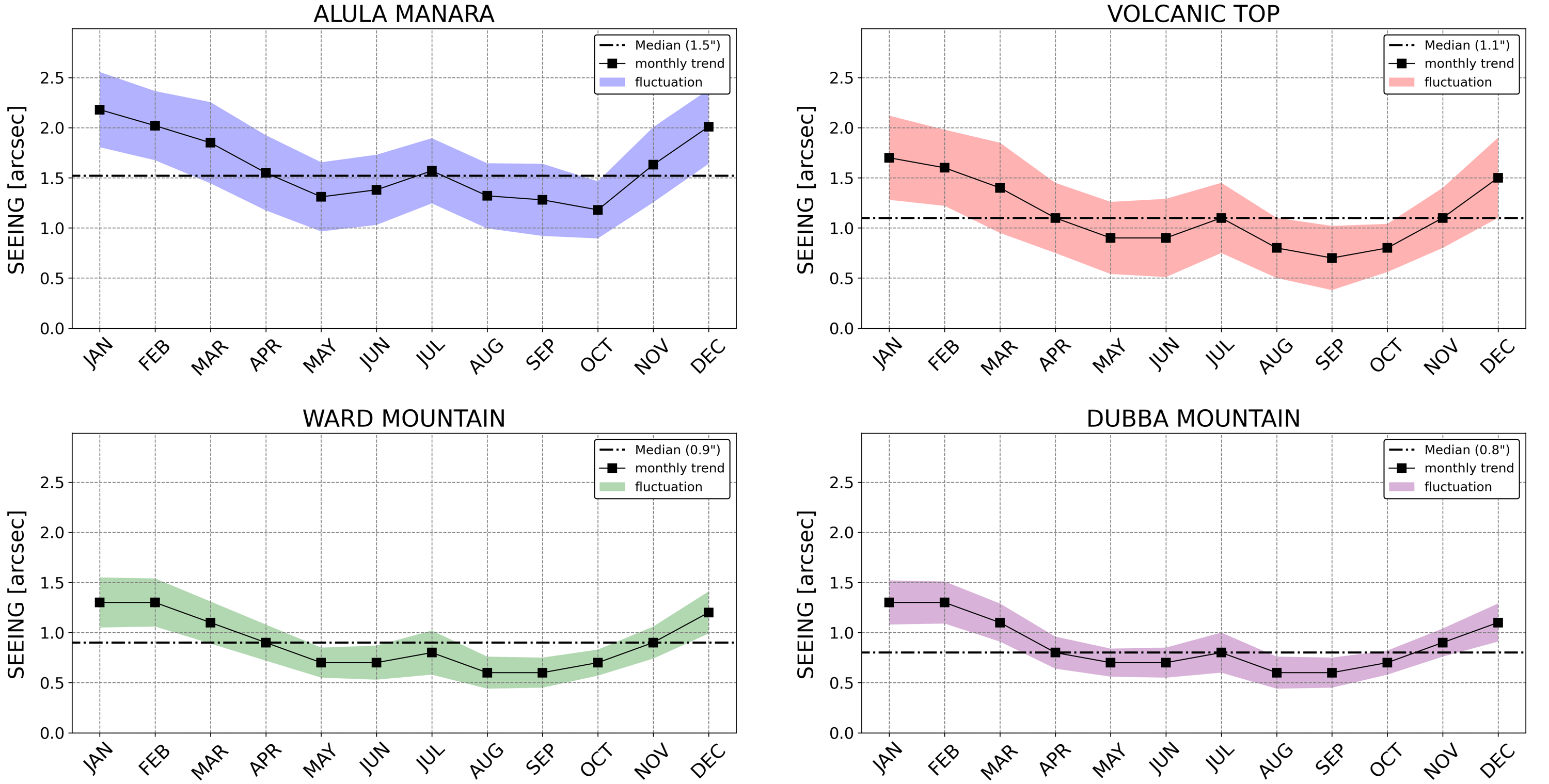}
\end{tabular}
\end{center}
    \caption{\label{fig:seeingmonthly} Monthly trend and fluctuation of the median seeing averaged through the years from the Gm model.}
\end{figure}
To identify seasonal changes we have calculated the nighttime median seeing $med(\epsilon)_{m,y}$ for every month $m$ in a specific year $y$. Afterwards, we get the time-frame monthly averages $\overline{med(\epsilon)_{m}}$ from equation \ref{eq:monthly_med}. Figure \ref{fig:seeingmonthly} shows the trend of the time-frame averages from the monthly median seeing for the full dataset retrieved from the $r_{0}$ data, including the fluctuation $\text{FL}_{m}^{\text{short}}(\epsilon)$ during a specific month (colored shade) calculated using equation \ref{eq:short_fluctuation}. We do see a clear seasonal tendency, but an inverse trend is found for the months of June/July. Furthermore, fluctuations are more pronounced at AlUla Manara and Volcanic Top. We suspect that, as lower and flatter sites compared to Ward Mountain and Dubba Mountain, they may exhibit more variability within the surface layer, extending into the Planetary Boundary Layer.\\
\begin{figure}
\begin{center}
\begin{tabular}{c}
\includegraphics[width=0.92\textwidth]{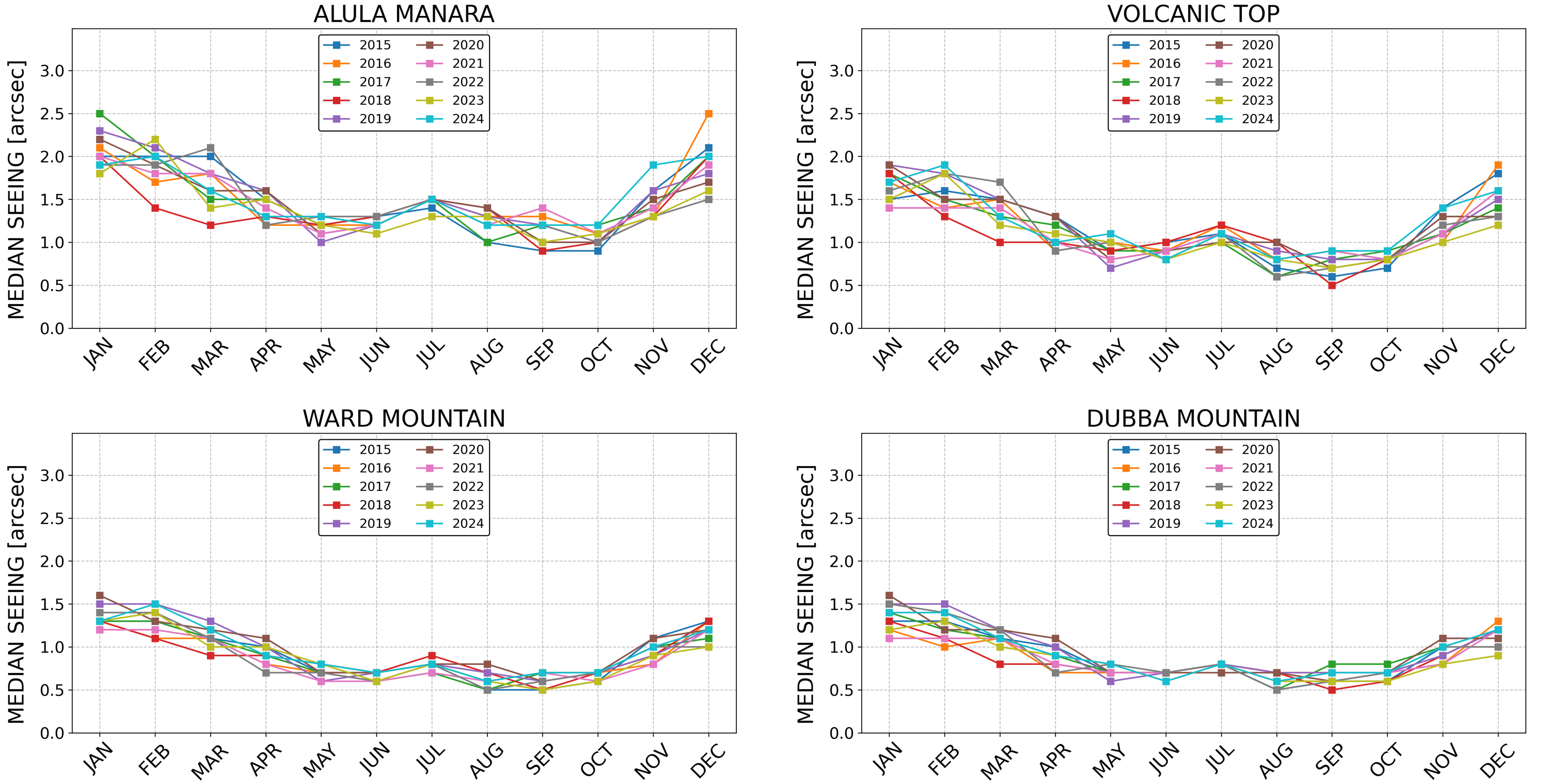}
\end{tabular}
\end{center}
    \caption{\label{fig:seeingall} Monthly trend of the median seeing for every year from the Gm model.}
\end{figure}
\begin{figure}
\begin{center}
\begin{tabular}{c}
\includegraphics[width=1.00\textwidth]{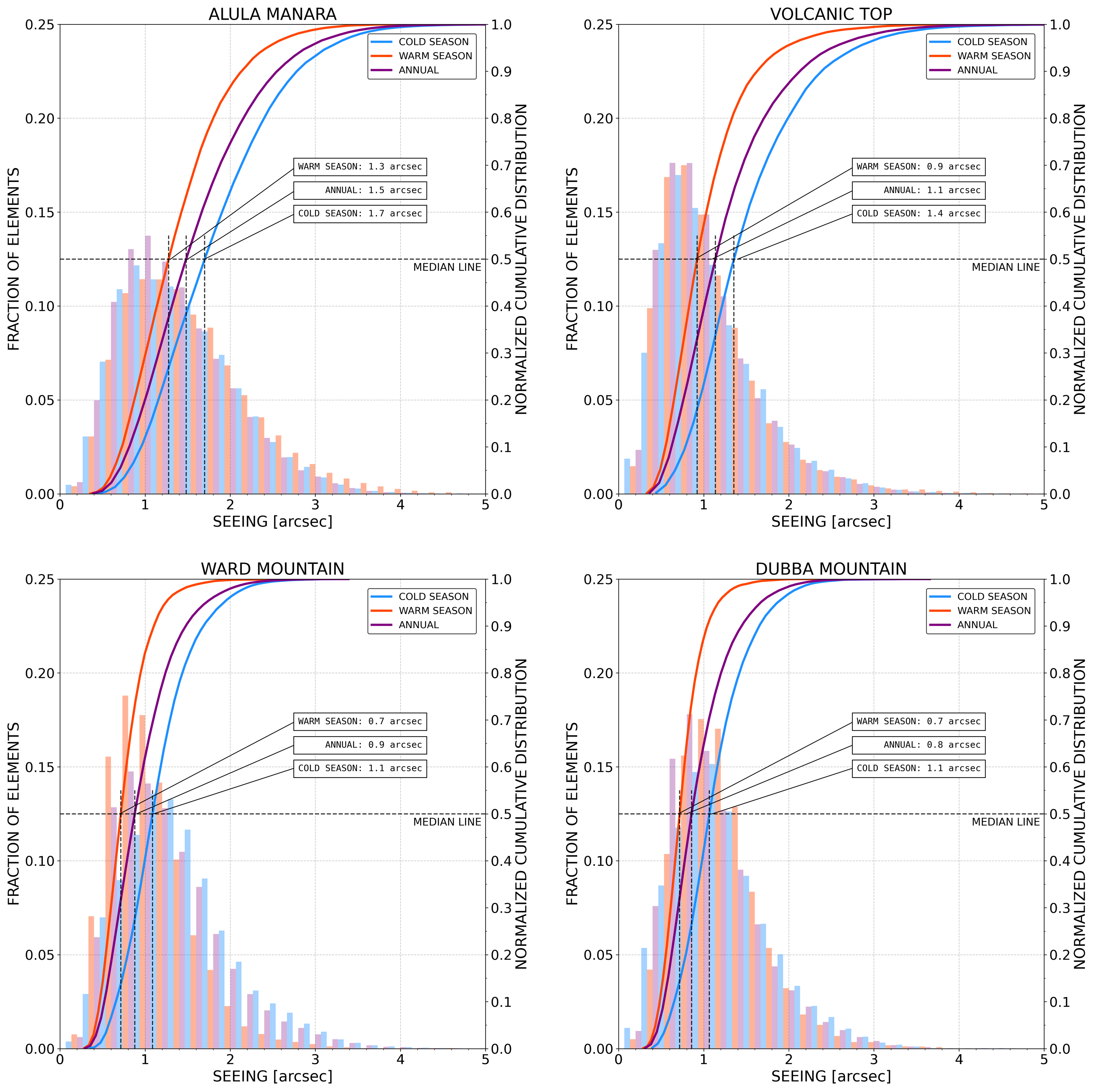}
\end{tabular}
\end{center}
\caption 
{\label{fig:seeingcumulative}
Cumulative distribution of the seeing from the Gm  model for all sites, with separation between the warm season and the cold season.} 
\end{figure}
Figure \ref{fig:seeingall} reports the seasonal trend for every year in the dataset. It's clear that the seasonal behavior of the monthly median seeing is well defined with better conditions typically occurring in the warm season. Figure \ref{fig:seeingcumulative} shows the cumulative distribution of the seeing from the Gm model for all sites. In the plots the warm and cold seasons have been separated and plotted side by side with the annual trend. This clearly shows how the values of the medians are significantly different when calculated in different periods. As expected, the estimated seeing is higher in the cold season than in the warm season. In Section \ref{sect:tgrad} we show weak seasonal trends in the day/night temperature gradient for all sites, so the calculated seeing could have a mild connection with that. In addition, a clear seasonal pattern in the wind direction is not found (see Section \ref{sect:wind}), so in principle the seasonal seeing trend should not be linked to the winds. All previous affirmations will be confirmed only after ASM data will be collected at AlUla Manara.\\
An important aspect to consider is the elevation range from which the $C_{n}^{2}$ profile is calculated by the model. Layers are sampled at finite altitudes that continuously vary in time with pressure level (see Table \ref{tab:metrics}). The first layer is typically set above the site location, thus the portion of atmosphere between the ground and the first layer remains unsensed. This implies a certain underestimation of the turbulent energy and thus the seeing. To avoid this circumstance a further layer can be added at the base of the modeled profile. Due to its finite nature, this further layer will partially cover an altitude range that will stay below the ground level, at negative altitudes. This implies a certain overestimation of the turbulent energy and thus the seeing.\\
\begin{table}
\caption{Average elevation of the first layer for each site together with the average elevation of the layer added at the base of the model ($first – 1$). The thickness represents the limits where the average altitude of the layers is calculated.}
\label{tab:layers}
\begin{center}       
\small
\renewcommand{\arraystretch}{1.2}
\begin{tabular}{|>{\columncolor{gray!15}}c|c|c|c|c|} 
\hline
 \cellcolor{white} & \textbf{Summit} & \textbf{Layer} & \textbf{Altitude}   & \textbf{Thickness} \\
 \cellcolor{white} & [m.a.s.l.]      &                & [m.a.s.l.]          & [m.a.s.l.] \\
\hline
\noalign{\vskip 3pt}
\hline
 \textbf{AlUla} & 1209 & \multicolumn{1}{r|}{$first$}   & 1381 & 1256.8 $\leftrightarrow$ 1506.5 \\
 \textbf{Manara} &     & \multicolumn{1}{r|}{$first-1$} & 1136 & 1014.9 $\leftrightarrow$ 1256.8 \\
\hline
\noalign{\vskip 3pt}
\hline
 \textbf{Volcanic} & 1807 & \multicolumn{1}{r|}{$first$}   & 1894 & 1762.3 $\leftrightarrow$ 2024.9 \\
 \textbf{Top}      &      & \multicolumn{1}{r|}{$first-1$} & 1634 & 1506.3 $\leftrightarrow$ 1762.3 \\
\hline
\noalign{\vskip 3pt}
\hline
 \textbf{Ward}     & 2052 & \multicolumn{1}{r|}{$first$}   & 2161 & 2026.3 $\leftrightarrow$ 2296.1 \\
 \textbf{Mountain} &      & \multicolumn{1}{r|}{$first-1$} & 1895 & 1763.4 $\leftrightarrow$ 2026.3 \\
\hline
\noalign{\vskip 3pt}
\hline
 \textbf{Dubba}    & 2146 & \multicolumn{1}{r|}{$first$}   & 2157 & 2022.3 $\leftrightarrow$ 2290.8 \\
 \textbf{Mountain} &      & \multicolumn{1}{r|}{$first-1$} & 1891 & 1760.5 $\leftrightarrow$ 2022.3 \\
\hline
\end{tabular}
\end{center}
\end{table}
\begin{table}
\caption{Median seeing in [arcsec] from the Gm model in different seasons with the standard $C_{n}^{2}$ profile and adding one more layer at the base of the model.} 
\label{tab:newseeing}
\begin{center}       
\small
\renewcommand{\arraystretch}{1.3}
\begin{tabular}{|>{\columncolor{gray!15}}c|r|c|c|c|} 
\hline
 \cellcolor{white}  & & \textbf{Annual} & \textbf{Warm}   & \textbf{Cold}\\
 \cellcolor{white}  & &        & \textbf{Season} & \textbf{Season}\\
\hline
\noalign{\vskip 2pt}
\hline
\textbf{AlUla}  & Standard        & 1.5 & 1.3  & 1.7\\
\textbf{Manara} &  One More Layer & 1.6 & 1.4  & 1.9\\ 
\hline
\noalign{\vskip 2pt}
\hline
\textbf{Volcanic} &  Standard       & 1.1 & 0.9  & 1.4\\
\textbf{Top}      &  One More Layer & 1.4 & 1.2  & 1.6\\ 
\hline
\noalign{\vskip 2pt}
\hline
\textbf{Ward}     &  Standard       & 0.9 & 0.7  & 1.1\\
\textbf{Mountain} &  One More Layer & 1.0 & 0.8  & 1.2\\ 
\hline
\noalign{\vskip 2pt}
\hline
\textbf{Dubba}    &  Standard       & 0.8 & 0.7  & 1.1\\
\textbf{Mountain} &  One More Layer & 0.9 & 0.8  & 1.1\\
\hline
\end{tabular}
\renewcommand{\arraystretch}{1}
\end{center}
\end{table}
Table \ref{tab:layers} shows the average elevation of the first layer for each site together with the average elevation of the layer added at the base of the model ($first – 1$). In the table, the thickness represents the limits where the average of the $first$ and the $first – 1$ layers altitudes were calculated. Thickness is presented this way because the $C_{n}^{2}(h)$ is estimated between two layers as it depends on vertical gradients of potential temperature $(\partial \theta/\partial z)$ and wind $(\partial u/\partial z, \partial v/\partial z)$, therefore the height of the $C_{n}^{2}$ is the average between two limits which are the heights of each layer in that moment given the instantaneous pressure level.\\
The increase of the estimated seeing with the addition of a further layer is reported in Table \ref{tab:newseeing}. From the considerations above, and interpolating the information from Table \ref{tab:layers} and Table \ref{tab:newseeing}, we may argue that the most accurate median seeing value could stay in between the two values, with the exception of Dubba Mountain, were the first layer almost match with the summit elevation. Nevertheless, the surface layer turbulence is probably not optimally estimated due to the lack of local measurements that could have facilitated proper calibration of the models. However, the increase of the estimated seeing with the addition of a further layer is less than 10\% for all sites, and probably contained withing the intrinsic model accuracy, so the direction that this analysis has taken is still consistent.
\section{Ground temperature analysis}\label{sect:t}
Ultraviolet radiation is absorbed by the atmosphere at a rate of about 5\%, while the atmosphere remains relatively transparent in the visible range. The radiation absorption is followed by a black body emission at 280 K having a peak at 12-13 $\mu$m (\citenum{Stephens1978}). This emission is absorbed by the atmospheric gases (prevalently in the H$_{2}$O and CO$_{2}$ bands) and generates an increase of temperature in the troposphere (\citenum{Petty2006}). This explains why the temperature decreases in altitude from sea level. The radiation absorption takes place in daytime when the ground receives solar radiation. This heat will be released during evening twilight, generating instability in the circulation of air in the boundary layer close to the ground. The day/night temperature gradient ($\Delta t$) contributes to the development of boundary layer turbulence, particularly during transition periods such as sunset and sunrise when temperature gradients are most pronounced.\\
The analysis of ground temperatures is performed by analyzing the surface-level 2 m temperature variable $t2m$ that represents the temperature at 2 m above ground for a given location. This parameter has units of kelvin [K] and has been converted to degrees Celsius [\textdegree C] by subtracting 273.15.
\begin{figure}
\begin{center}
\begin{tabular}{c}
\includegraphics[width=0.75\textwidth]{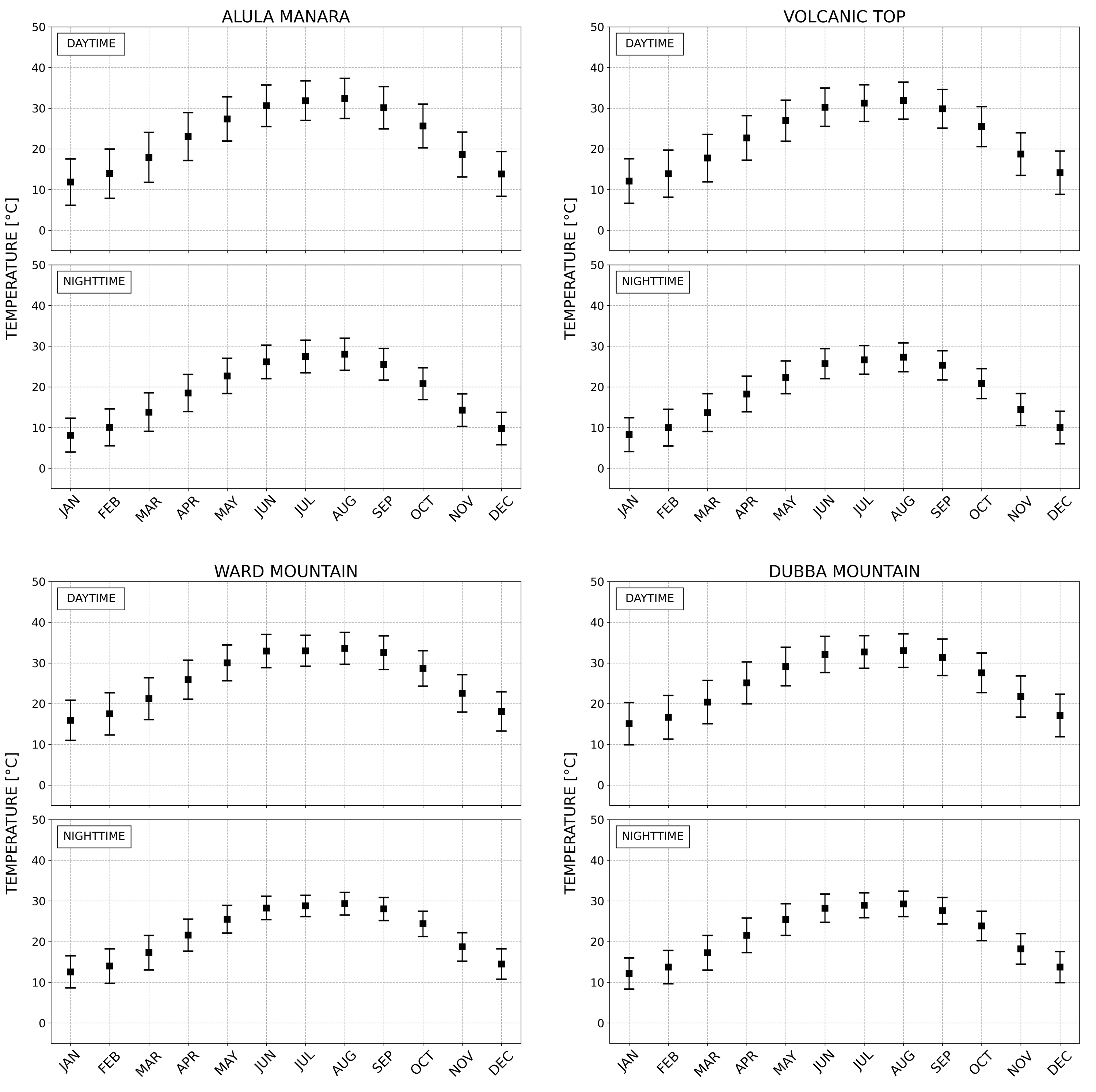}
\end{tabular}
\end{center}
    \caption{\label{fig:temp} Daytime and nighttime 2 m temperature seasonal trend in terms of monthly averages calculated through the years ($\overline{avg(t2m)_{m}} \pm \sigma_{avg,y}$).}
\end{figure}
\begin{table}
    \centering
    \caption{\label{tab:stat_tempe} Statistics of the 2 m temperature in [\textdegree C] for all sites. Values are reported for daytime and nighttime in the Saudi Arabia warm and cold seasons.}
    \renewcommand{\arraystretch}{1.2}
    \setlength{\tabcolsep}{5pt} % Adjust spacing
    \arrayrulecolor{black} % Ensure table lines are black
    \small
    \begin{tabular}{|>{\columncolor{gray!15}}c|r|rrrrr|rrrrr|}
        \hline
        \cellcolor{white} & \multicolumn{1}{|c|}{} & \multicolumn{5}{c}{\textbf{Warm Season}} & \multicolumn{5}{|c|}{\textbf{Cold Season}} \\ 
        \cline{3-12}
        \cellcolor{white} & \multicolumn{1}{|c|}{} & Min & 25\% & Avg & 75\% & Max & Min & 25\% & Avg & 75\% & Max \\ 
        \hline
        \noalign{\vskip 2pt}
        \hline
        \textbf{AlUla} & Daytime & 9.6 & 28.8 & 30.9 & 33.5 & 41.8 & $-1.3$ & 14.7 & 18.2 & 20.7 & 37.2 \\ 
        \textbf{Manara} & Nighttime & 3.7 & 21.6 & 23.6 & 26.2 & 39.6 & $-5.6$ &  8.4 & 11.8 & 14.1 & 34.1\\
        \hline
        \noalign{\vskip 2pt}
        \hline
        \textbf{Volcanic} & Daytime & 7.7 & 27.7 & 29.7 & 32.1 & 40.6 & $-0.3$ & 14.5 & 17.8 & 20.2 & 36.5 \\ 
        \textbf{Top} & Nighttime & 3.0 & 21.8 & 23.7 & 26.1 & 39.2 & $-5.3$ &  8.9 & 12.2 & 14.6 & 34.0\\ 
        \hline
        \noalign{\vskip 2pt}
        \hline
        \textbf{Ward} & Daytime & 11.8 & 29.8 & 31.1 & 33.0 & 44.3 & 3.9 & 18.0 & 21.1 & 23.5 & 39.3 \\ 
        \textbf{Mountain} & Nighttime &  7.5 & 26.0 & 27.4 & 29.4 & 42.6 & 0.2 & 13.2 & 16.5 & 19.0 & 37.6\\
        \hline
        \noalign{\vskip 2pt}
        \hline
        \textbf{Dubba} & Daytime & 13.2 & 30.0 & 33.2 & 43.0 & 42.5 &    5.8 & 17.8 & 21.0 & 23.5 & 38.0 \\ 
        \textbf{Mountain} & Nighttime &  8.0 & 24.9 & 28.1 & 28.8 & 41.2 & $-0.1$ & 11.8 & 15.2 & 17.8 & 36.9\\
        \hline
    \end{tabular}
\end{table}
\subsection{Absolute ground temperature}\label{sect:tabs}
To identify seasonal changes, data has been averaged for every month $m$ through the years $y$ as monthly averages $avg(t2m)_{m,y}$. Afterwards, the monthly averages of the same month in different years have been averaged using equation \ref{eq:monthly_avg}. In Figure \ref{fig:temp} the 2 m temperature seasonal trends in terms of $\overline{avg(t2m)_{m}} \pm \sigma_{avg,y}$ for all sites are shown for daytime and nighttime data, while Table \ref{tab:stat_tempe} reports the statistics of the 2 m temperature in [\textdegree C] for all sites for the whole time-frame. In the table the results are divided for the Saudi Arabia warm and cold seasons, and in each season for daytime and nighttime.\\
The average temperatures are very similar between the sites, with slightly higher estimations for Ward Mountain and Dubba Mountain. The climate of the sites is characterized by very hot to warm seasons with average temperatures around 30\textdegree C, and extreme events of maximums above 40\textdegree C in both daytime and nighttime. Average temperatures during the night can oscillate between 25 to 28\textdegree C.\\
In the cold season the temperatures have a mild drop with averages of 18-21\textdegree C in daytime and 12 to 17\textdegree C in nighttime, with sporadic extreme events of negative temperatures at night. Temperature maximums can still reach values as high as 39\textdegree C in daytime and about 37\textdegree C during the night.
\begin{table}
\caption{Statistics of the day/time temperature gradient $\Delta t$ in [\textdegree C] for all sites.} 
\label{tab:Dt}
\begin{center}
\small
\begin{tabular}{|r|c|c|c|c|c|} 
\hline
\rule[-1ex]{0pt}{3.5ex}  & \textbf{min} & \textbf{25\%} & \textbf{average} & \textbf{75\%} & \textbf{max} \\
\hline\hline
\rule[-1ex]{0pt}{3.5ex} \textbf{AlUla Manara} & 2.9 & 11.5 & 12.9 $\pm$ 2.4 & 14.5 & 20.6 \\ 
\hline
\rule[-1ex]{0pt}{3.5ex} \textbf{Volcanic Top} & 3.1 & 10.9 & 12.2 $\pm$ 2.1 & 13.6 & 19.5 \\  
\hline
\rule[-1ex]{0pt}{3.5ex} \textbf{Ward Mountain} & 2.8 & 8.8 & 10.4 $\pm$ 2.3 & 12.1 & 18.5 \\ 
\hline
\rule[-1ex]{0pt}{3.5ex} \textbf{Dubba Mountain} & 2.1 & 9.8 & 11.3 $\pm$ 2.4 & 13.0 & 19.0 \\ 
\hline
\end{tabular}
\end{center}
\end{table}
\subsection{Day/night temperature gradient}\label{sect:tgrad}
\begin{figure}
\begin{center}
\begin{tabular}{c}
\includegraphics[width=0.98\textwidth]{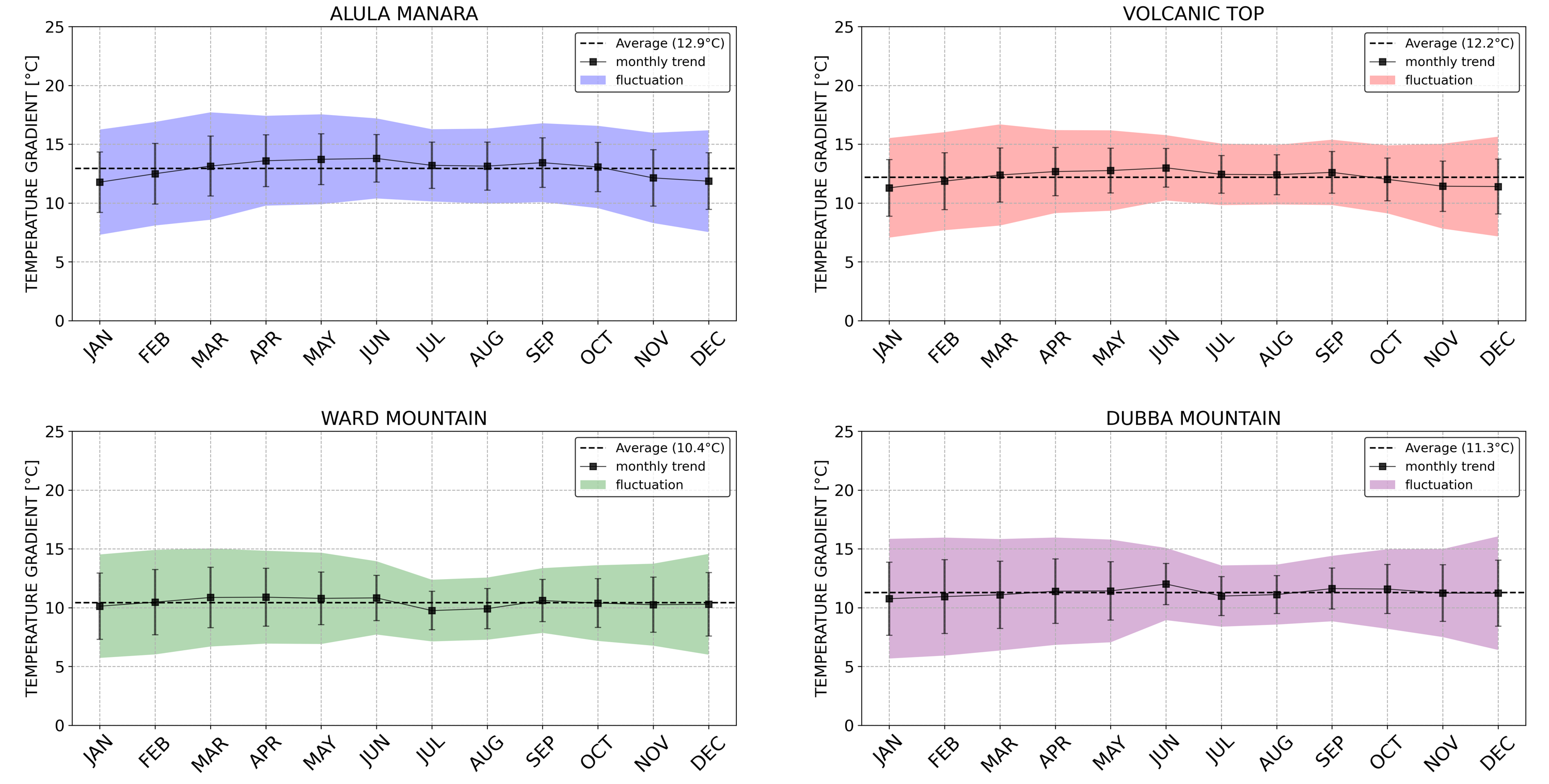}
\end{tabular}
\end{center}
    \caption{\label{fig:Dt} Day/night temperature gradient $\Delta t$ seasonal trend as monthly averages calculated through the years for the full dataset ($\overline{avg(\Delta t)_{m}} \pm \sigma_{avg,y}$).}
\end{figure}

\begin{figure}
\begin{center}
\begin{tabular}{c}
\includegraphics[width=0.70\textwidth]{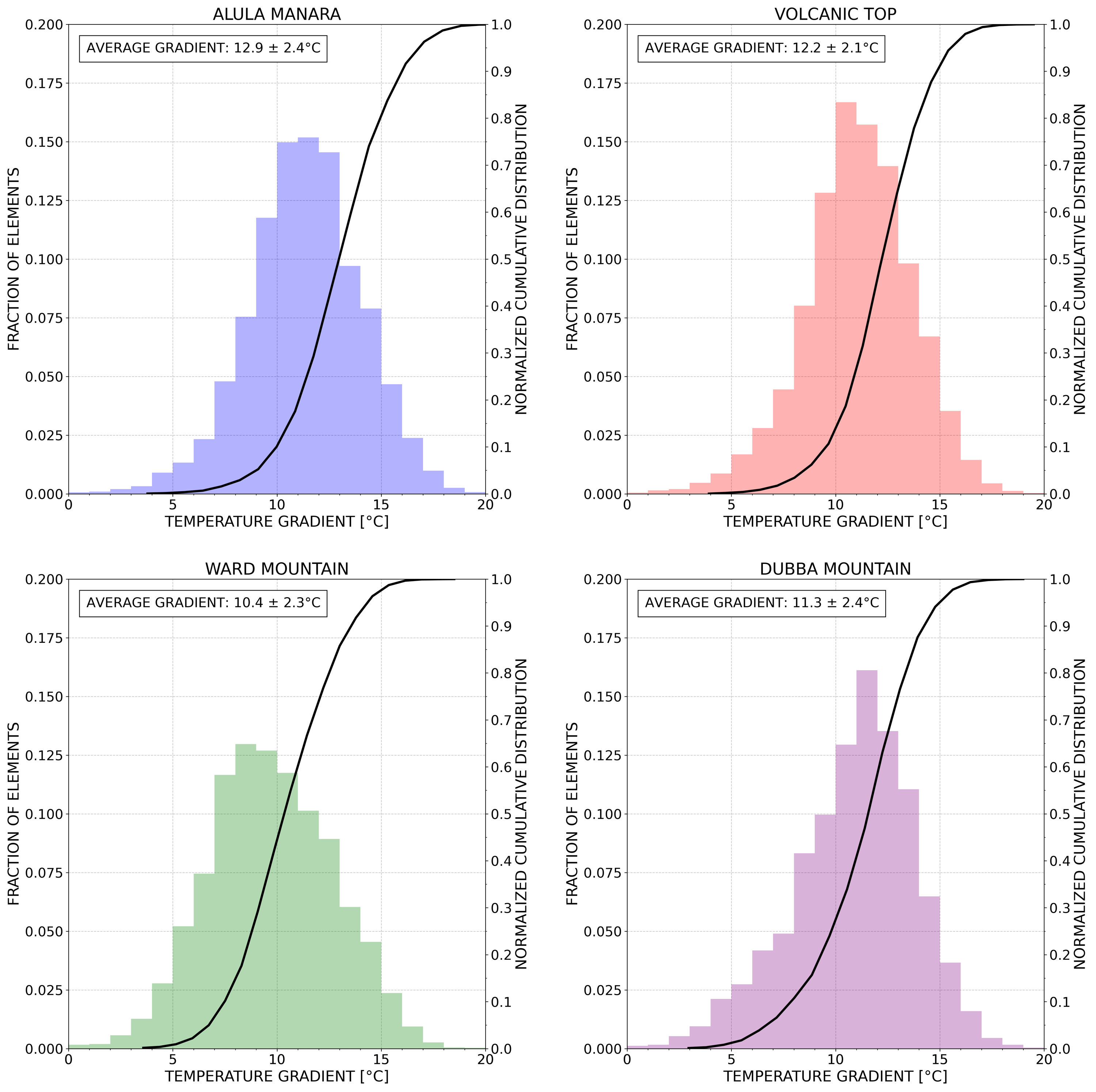}
\end{tabular}
\end{center}
    \caption{\label{fig:DtCumul} Distribution of the daily day/night temperature gradient $\Delta t$ for all sites between 1990 and 2024.}
\end{figure}

The day/night temperature gradient $\Delta t$ is defined as the difference between the daytime maximum and the nighttime minimum among 24 hours. Table \ref{tab:Dt} reports the statistics of $\Delta t$ in [\textdegree C] for all sites retrieved from the full dataset between 1990 and 2024. To identify seasonal changes, the daily $\Delta t$ have been averaged during a specific month, and then averaged for the same month through the years using equation \ref{eq:monthly_avg}. Results are reported in Figure \ref{fig:Dt} where we have included the fluctuation $\text{FL}_{m}^{\text{long}}(\Delta t)$ (colored shade) as calculated from equation \ref{eq:long_fluctuation}. In the figure the error bars indicate that the variability in a single month is relatively high compared to the month-to-month variation, suggesting a weak $\Delta t$ the seasonal trend.\\
In Figure \ref{fig:DtCumul} we show the cumulative distribution of the daily day/time temperature gradient for all sites from the full dataset. The average gradients stand around 10 to 13\textdegree C. Gradients above 15\textdegree C are reached less than 5\% of the time for all sites, while very low gradients (negligible difference between daytime maximum and nighttime minimum temperatures, i.e. 3\textdegree C) are unusual with frequencies below 2\%.
\begin{figure}
\begin{center}
\begin{tabular}{c}
\includegraphics[width=1.00\textwidth]{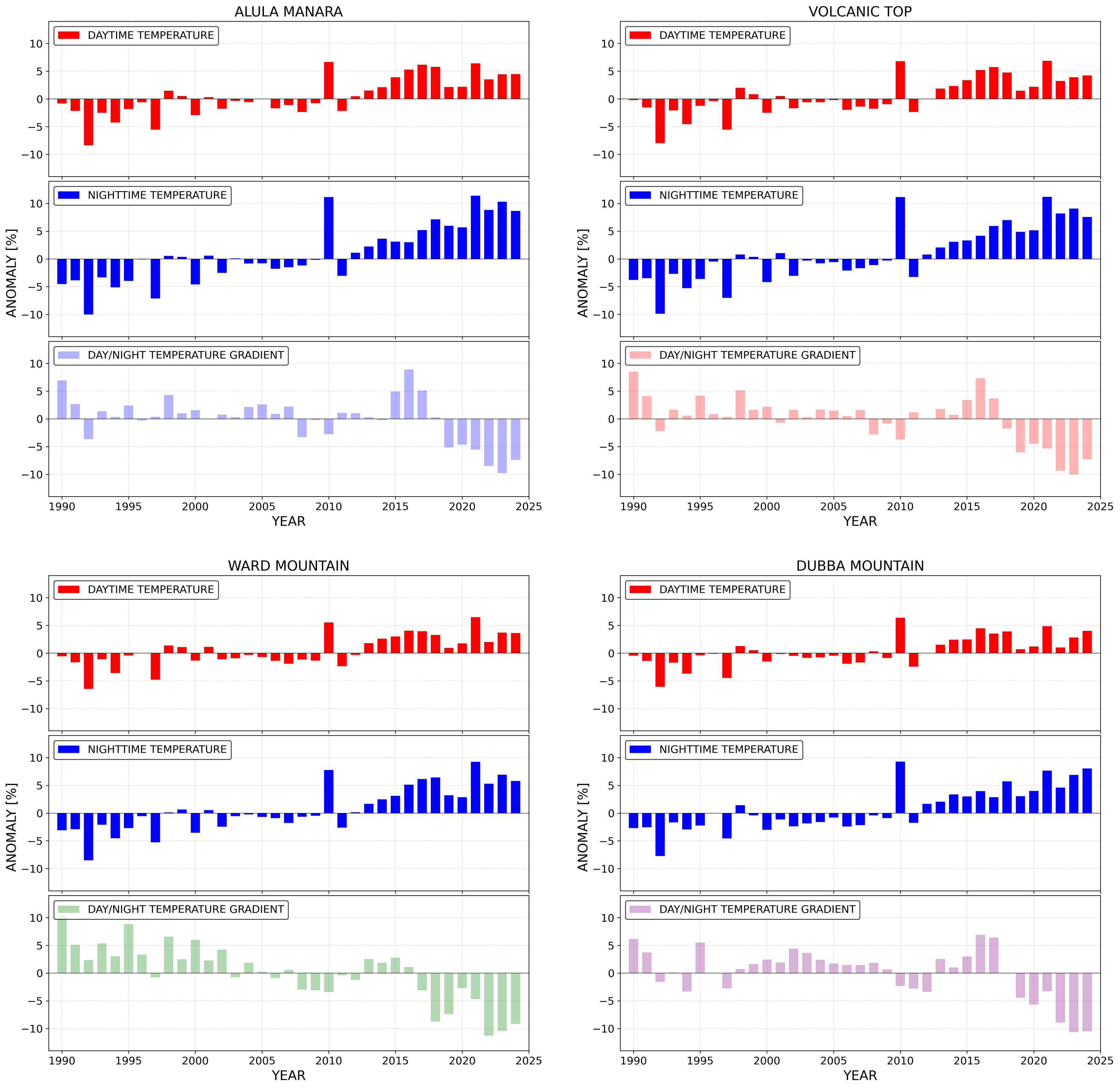}
\end{tabular}
\end{center}
    \caption{\label{fig:t_anom} Anomalies for the daytime and nighttime temperatures, and for the day/night temperature gradient.}
\end{figure}
\subsection{Temperature anomalies}\label{sect:tanom}
The climate anomaly is the difference between an observed climate variable and a reference value based on a long-term average (\citenum{lamb1972climate}). It allows to detect trends and variations in climate over time to identify long-term trends. The standard 30-year period used to calculate climate anomalies is called "the climate normal" and is defined by the World Meteorological Organization (WMO). The most recent standard period is 1991--2020 (\citenum{arguez2011definition}), but older reference periods like 1961--1990 or 1981--2010 are also commonly used. In our case we have used the last updated climate normal (1991--2020) to reflect the long-term climate trends without being outdated.\\
To calculate the temperature anomalies we have averaged the hourly \textit{t2m} temperatures in a single year to get the yearly averages $\overline{t_{y}}$, separately for data in the daytime and nighttime time ranges. Then, the temperature normal $t_{norm}$ is calculated averaging the $t2m$ over the 30 years period 1991--2020, respectively for data in the daytime and nighttime time ranges. Finally, the temperature anomalies are calculated as the relative difference between the averaged yearly daytime or nighttime temperature $\overline{t_{y}}$ and the 30-year daytime or nighttime normals $t_{norm}$:
\vspace{0.5em}
\begin{equation}
\label{eq:anomaly}
t_{anom} = \frac{\overline{t_{y}} - t_{norm}}{t_{norm}},
\end{equation}
For what concerns the anomalies of the day/night temperature gradient, the daily gradients have been averaged among every single year, providing the yearly averages $\overline{\Delta t_{y}}$, while the normal $\Delta t_{norm}$ has been calculated averaging the daily $\Delta t$ over the 30 years period 1991--2020. So, the day/night temperature gradient anomalies are calculated as:
\vspace{0.5em}
\begin{equation}
\label{eq:anomalydt}
\Delta t_{anom} = \frac{\overline{\Delta t_{y}} - \Delta t_{norm}}{\Delta t_{norm}},
\end{equation}
In Figure \ref{fig:t_anom} we plot the calculated anomalies for the temperatures and the day/night temperature gradient. The plot indicates an increase of the temperatures in the last decade. Nevertheless, the increase is somewhat more pronounced for the nighttime temperatures with the consequence that the day/night temperature gradients $\Delta t$ is affected by a progressive decrease of up to 10\% of the averaged value.\\
The observed long-term decrease in the day/night temperature gradient may indicate a gradual thermal homogenization of the near-surface atmosphere. This diminishing gradient could reflect reduced radiative cooling during nighttime or broader climate-induced changes in the surface energy balance, both of which carry implications for the long-term thermal regime of the sites. This trend may influence the development of near-surface convective instability and, consequently, the strength and structure of surface layer turbulence. For these reasons, we consider this evolution particularly relevant and recommend that it be carefully monitored in the coming years.\\
\begin{table}
\caption{Wind speed statistics in [m/s] for all sites and dominant wind directions.} 
\label{tab:wsp}
\renewcommand{\arraystretch}{1.2}
\begin{center}       
\small
\begin{tabular}{|>{\columncolor{gray!15}}c|r|c|c|c|c|c|c|} 
\hline
\cellcolor{white} & & \textbf{25\%} & \textbf{average} & \textbf{75\%} & \textbf{max} & & $w_{dir}$\\  
\hline
\noalign{\vskip 3pt}
\hline
\textbf{AlUla} & Daytime   & 2.4 & 3.8 $\pm$ 2.0 & 4.8 & 15.1 & & WSW\\  
\textbf{Manara} &   Nighttime & 2.0 & 2.9 $\pm$ 1.4 & 3.6 & 13.3 & & W to N\\
\hline
\noalign{\vskip 3pt}
\hline
\textbf{Volcanic} &    Daytime   & 2.4 & 4.2 $\pm$ 2.2 & 5.7 & 15.0 & & WSW\\  
 \textbf{Top} &   Nighttime & 1.8 & 2.8 $\pm$ 1.5 & 3.4 & 13.4 & & WSW to NE\\
\hline
\noalign{\vskip 3pt}
\hline
 \textbf{Ward} &   Daytime   & 1.6 & 3.1 $\pm$ 1.6 & 4.4 & 9.1 & & W\\  
\textbf{Mountain} &    Nighttime & 1.2 & 1.9 $\pm$ 1.0 & 2.4 & 8.3 & & NE\\
\hline
\noalign{\vskip 3pt}
\hline
 \textbf{Dubba} &   Daytime   & 1.2 & 2.4 $\pm$ 1.3 & 3.5 & 7.5 & & WSW to W\\  
 \textbf{Mountain} &   Nighttime & 1.3 & 1.9 $\pm$ 0.8 & 2.5 & 7.2 & & NE to ESE\\  
\hline
\end{tabular}
\end{center}
\end{table}
\section{Wind pattern}\label{sect:wind}
\begin{figure}
\begin{center}
\begin{tabular}{c}
\includegraphics[height=0.95\textheight]{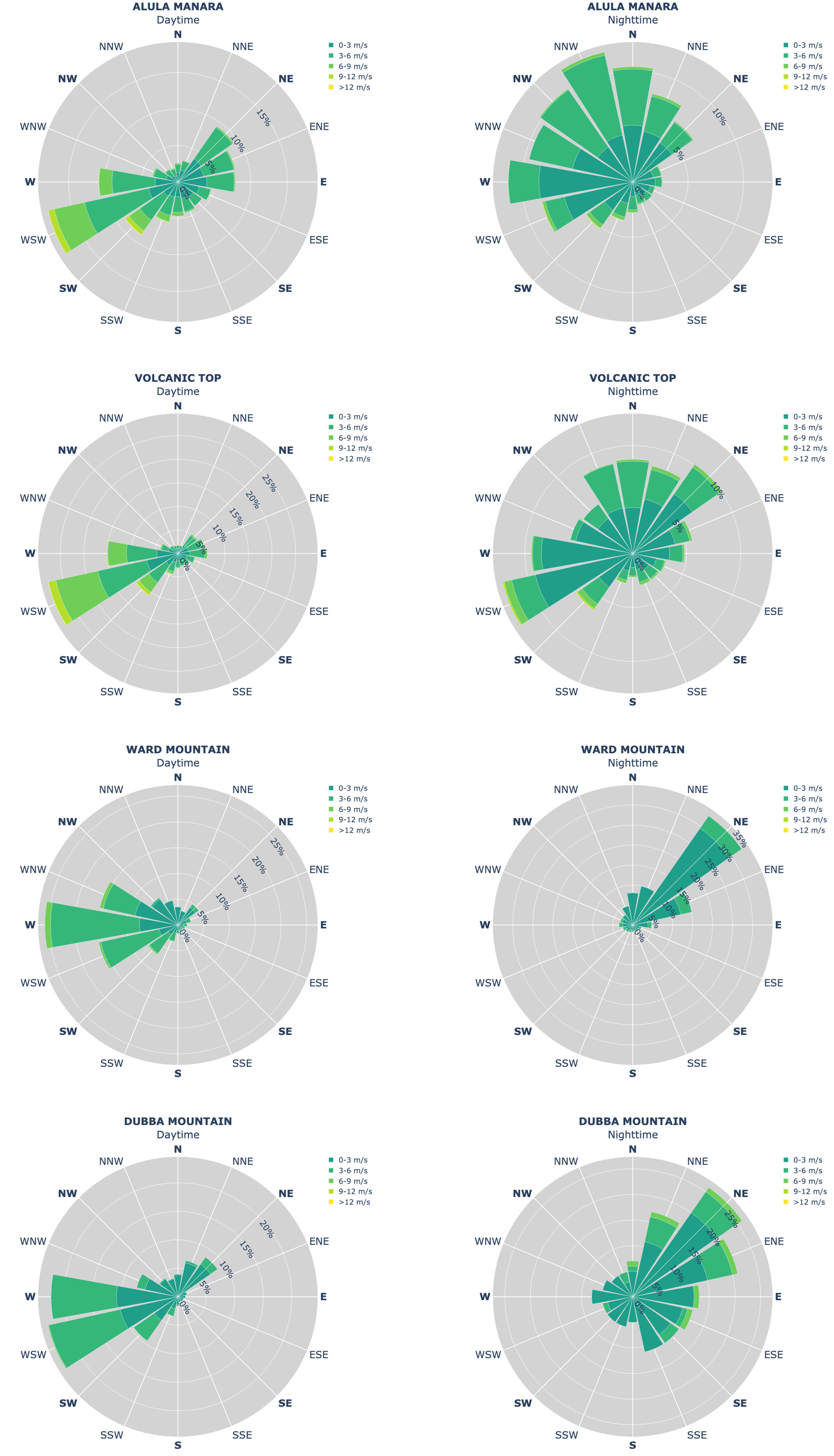}
\end{tabular}
\end{center}
    \caption{\label{fig:wdir} Wind direction patterns in daytime (left) and nighttime (right).}
\end{figure}

\begin{figure}
\begin{center}
\begin{tabular}{c}
\includegraphics[width=0.77\textwidth]{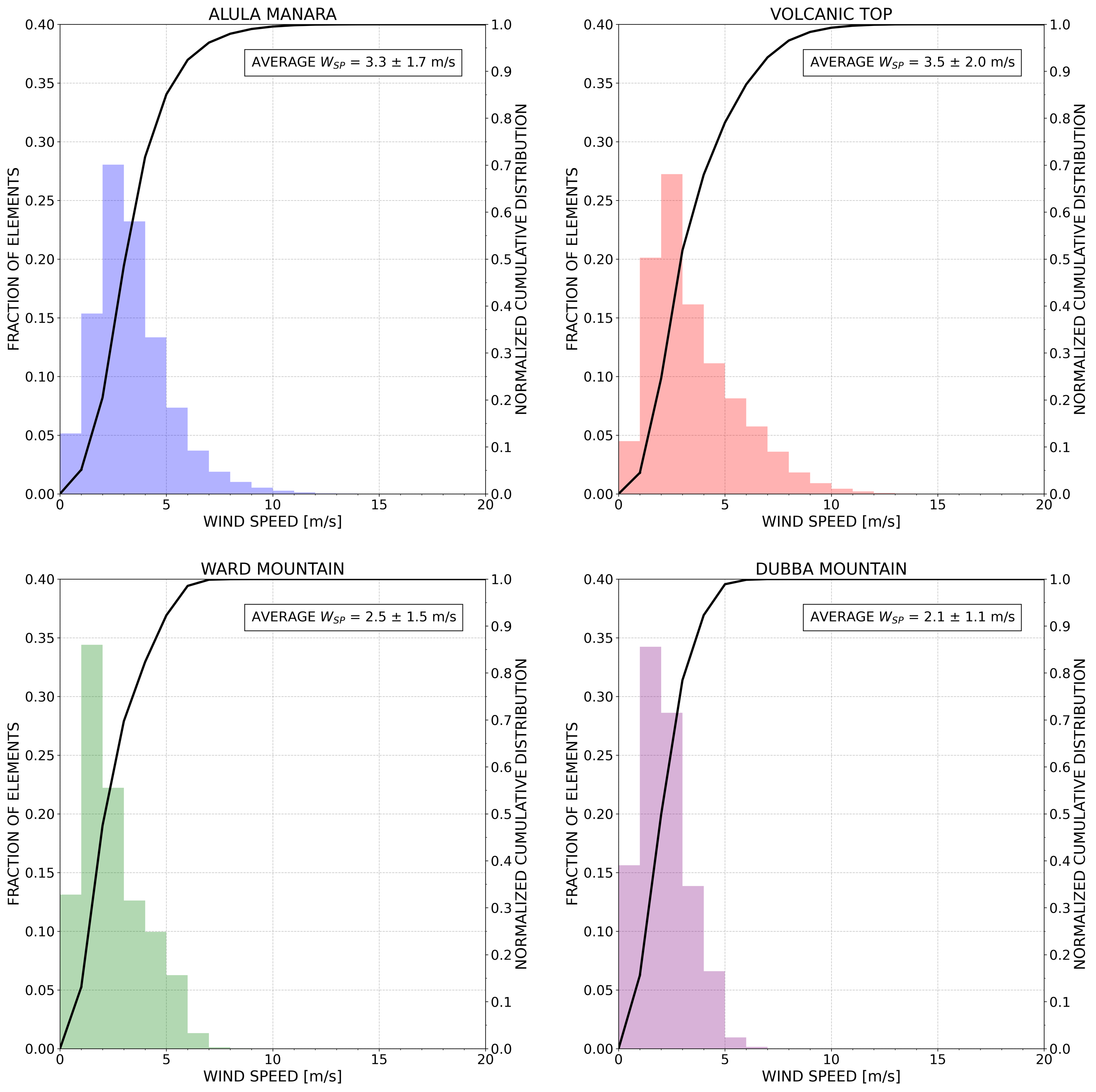}
\end{tabular}
\end{center}
    \caption{\label{fig:wspCumul} Distribution of the 24-hours wind speed for all sites between 1990 and 2024.}
\end{figure}

The wind pattern analysis is performed differentiating the daytime and nighttime behavior of the winds. A further distinction is made by separately considering the behavior of the winds during the warm and cold seasons. For the four sites, the following variables have been downloaded:
\begin{itemize}
\item \textbf{10m u-component of wind ($u10$)}. The eastward component of the 10 m wind. It is the horizontal speed of air moving towards the east, at a height of 10 m above the surface of the Earth, in [m/s].
\item \textbf{10m v-component of wind ($v10$)}. The northward component of the 10 m wind. It is the horizontal speed of air moving towards the north, at a height of 10 m above the surface of the Earth, in [m/s]. 
\end{itemize}
The $u10$ and $v10$ vectors can be combined to provide the estimated wind speed and direction of the horizontal 10 m wind from the atmospheric reanalysis.\\
The wind speed ($w_{sp}$) is calculated from the $u10$ and $v10$ scalars with the formula:
\begin{equation}
\label{eq:wsp}
w_{sp} = \sqrt{u10^{2} + v10^{2}}.
\end{equation}
The wind direction pattern ($w_{dir}$) can be retrieved combining the $u10$ and $v10$ vectors with the Python library \texttt{metpy.calc.wind\_direction} which automatically applies the following formula:
\begin{equation}
\label{eq:wdir}
w_{dir} = \mod{\left( 180 + \frac{180}{\pi}\atantwo(v10, u10), 360 \right)}.
\end{equation}
The outcomes of the analysis is reported in the Tables \ref{tab:wsp} and in the wind-roses plots of Figure \ref{fig:wdir}. The results include the whole analyzed period 1990-2024, without differentiating per year. We have found a remarkable difference in the wind direction patterns between daytime and nighttime for all sites (see Figure \ref{fig:wdir}). Conversely, we have not found evidences of seasonal changes, meaning that the differences between daytime and nighttime wind patterns are not season dependent.\\
The wind speed statistics is indicative of low winds all year around for all sites (3 to 4 m/s), with peaks never above 16 m/s for AlUla Manara and Volcanic Top, and 8 to 9 m/s for Ward Mountain and Dubba Mountain. To further support these affirmations, in Figure \ref{fig:wspCumul} we report the distribution of the wind speed for the full dataset, without distinguishing between daytime and nighttime, confirming the low wind speed pattern characterizing all sites.\\ 
\section{Cloud coverage and fraction of hours with clear sky}\label{sect:tcc}
The evaluation of remote sensed cloud coverage can give hints about the expected fraction of hours with clear sky above a given location. This analysis is performed with the use of the \textit{total cloud cover} variable ($tcc$). The total cloud cover corresponds to the fraction of the sky covered by clouds and in measured in [\%]. As described in Section {\ref{sect:intro}}, AlUla Manara is a large wide plateau, with no prominent peak around. For this reason we have decided to use the total cloud cover instead of layered cloud coverage (low-, mid-, high-) to fix a more restrictive parameter and avoid overestimating the number of clear hours. With this choice we expect some overestimation of the cloud coverage for Ward Mountain and Dubba Mountain, which are prominent peaks instead.\\
This is a simple and straightforward analysis, where data have been averaged for every month through the years in order to identify seasonal changes  (see equation \ref{eq:monthly_avg}). Similarly to what we have done for the day/night temperature gradient, also in this case we have calculated fluctuations corresponding to the averaged span between maximum and minimum coverage reached in a specific month using equation \ref{eq:long_fluctuation}.\\
\begin{table}
\caption{Average total cloud cover in terms of fraction of the sky covered by clouds [\%] and percentage of hours with clear sky (total cloud cover less than 10\%) in daytime and nighttime on annual basis, and for the warm and cold seasons.}
\label{tab:tcccln}
\renewcommand{\arraystretch}{1.2}
\begin{center}  
\small
\begin{tabular}{|>{\columncolor{gray!15}}c|r|c|c|c|c|c|c|c|}
\hline
\cellcolor{white} & & \multicolumn{3}{c|}{\textbf{Total Cloud Cover}} & & \multicolumn{3}{c|}{\textbf{Percentage of clear hours}}\\
\hline
\cellcolor{white} & & Annual & Warm & Cold & & Annual & Warm & Cold\\  
\cellcolor{white} & &      & Season & Season & &  & Season & Season\\  
\hline
\noalign{\vskip 3pt}
\hline
 \textbf{AlUla} & Daytime   & 17.9 & 13.9 & 21.8 & & 73.2 & 78.4 & 68.9\\  
 \textbf{Manara} &  Nighttime & 14.6 & 8.6 & 20.5 & & 79.4 & 87.4 & 71.4\\
\hline
\noalign{\vskip 3pt}
\hline
 \textbf{Volcanic} &  Daytime   & 18.7 & 15.1 & 22.4 & & 72.3 & 76.9 & 67.7\\  
 \textbf{Top} &  Nighttime & 14.9 & 8.9 & 20.9 & & 78.8 & 86.7 & 70.7\\
\hline
\noalign{\vskip 3pt}
\hline
 \textbf{Ward} &  Daytime   & 16.5 & 12.3 & 20.8 & & 75.5 & 80.9 & 70.0\\  
 \textbf{Mountain} &  Nighttime & 14.2 & 9.5 & 18.9 & & 79.2 & 85.3 & 73.0\\
\hline
\noalign{\vskip 3pt}
\hline
 \textbf{Dubba} &  Daytime   & 13.1 & 7.8 & 18.5 & & 81.3 & 88.6 & 74.0\\  
 \textbf{Mountain} &  Nighttime & 12.0 & 6.5 & 17.6 & & 82.8 & 90.6 & 75.1\\
\hline
\end{tabular}
\end{center}
\end{table}
\begin{figure}
\begin{center}
\begin{tabular}{c}
\includegraphics[height=0.95\textheight]{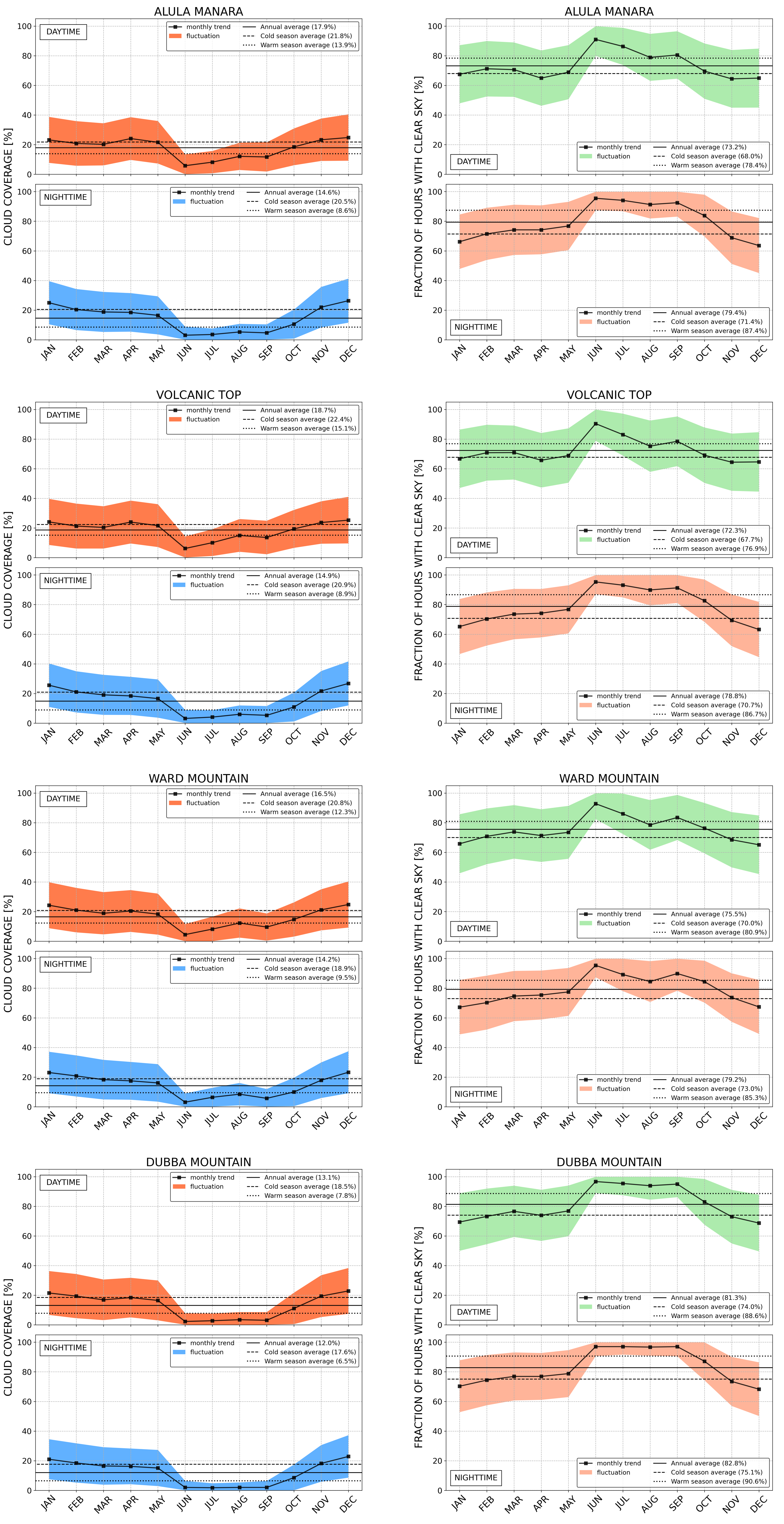}
\end{tabular}
\end{center}
    \caption{\label{fig:tcccln} Seasonal trends of the total cloud cover (left) and the percentage of clear hours (right) for all sites.}
\end{figure}
Results are reported in Table \ref{tab:tcccln} where the average total cloud cover in daytime and nighttime for the warm season, cold season and on an annual basis is shown together with relative results of the percentage of hours with clear sky.\\
\\
The criteria to define the sky as \textit{clear} follows the rule adopted by the European Southern Observatory, that is when the cloud coverage is less than 10\% (\textit{clear} to \textit{thin} conditions). This definition does not imply the complete absence of clouds and therefore does not necessarily correspond to zero total cloud cover in a binary classification.\\
\\
In Figure \ref{fig:tcccln} seasonal trends of the total cloud cover and the percentage of clear hours for all sites are presented. In the plots the fluctuations correspond to the averaged span between maximum and minimum reached in a specific month (equation \ref{eq:long_fluctuation}). The total cloud cover is fairly low, with higher presence in the cold season. Comparing daytime and nighttime data we can see that at nighttime conditions are generally better than in daytime. The percentage of hours with clear sky are above 65\% in the cold season, and above 75\% in the warm season for all sites, with significantly better conditions during nighttime.
\section{Prediction of the column of Water Vapor}\label{sect:pwv}
\begin{figure}
\begin{center}
\begin{tabular}{c}
\includegraphics[height=9cm]{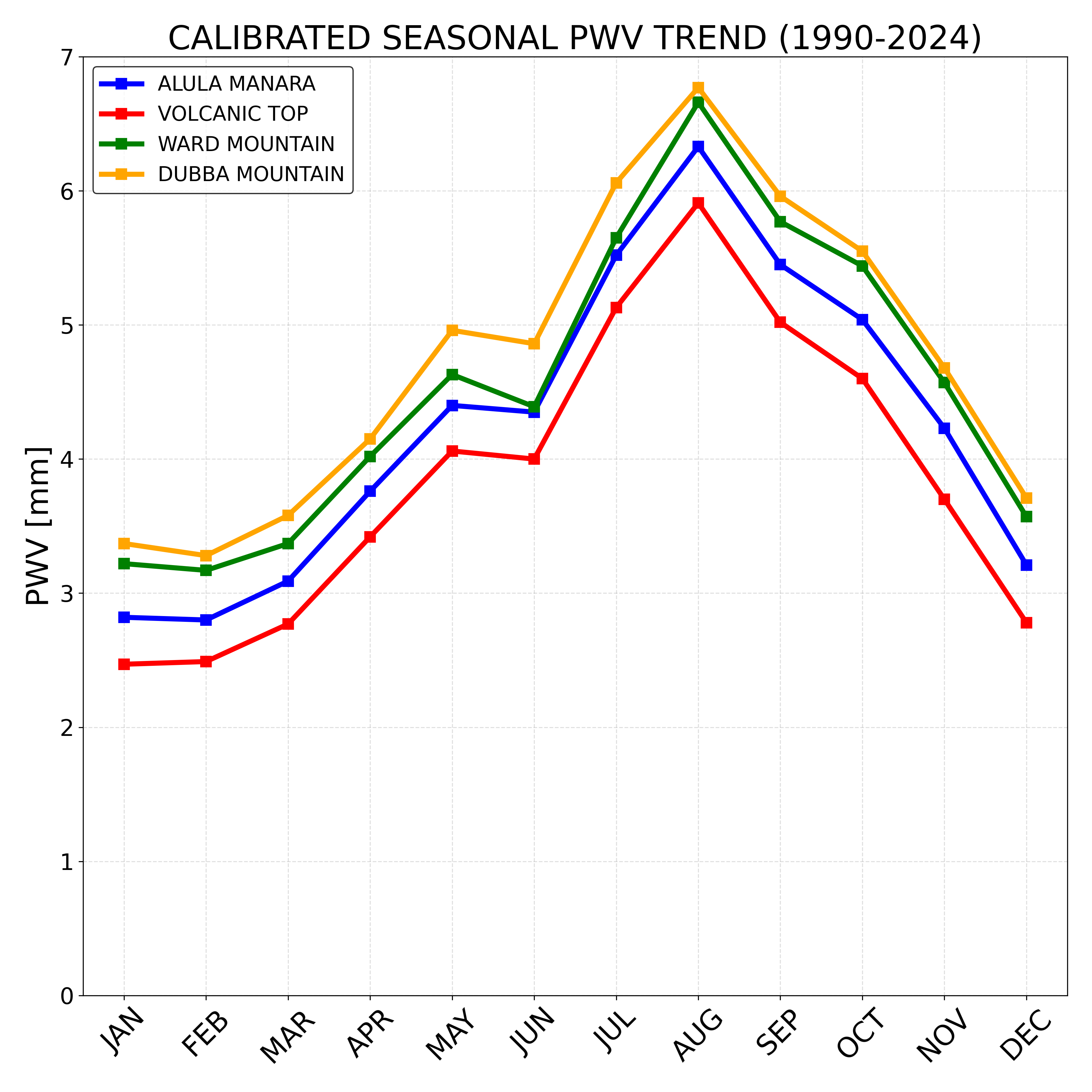}
\end{tabular}
\end{center}
\caption 
{\label{fig:pwvseason} Median seasonal PWV in [mm] averaged through the years from calibrated data for the Saudi Arabia sites.} 
\end{figure}

The column of Water Vapor above a location is expressed in terms of Integrated Water Vapor (IWV, in [kg/m$^2$]), also known as Precipitable Water Vapor (PWV, in [mm]). It represents the total content of water vapor in a 1 m$^2$ base column of atmosphere above a specific location. Water vapor absorbs and scatters electromagnetic radiation in the infrared and microwave wavelengths, reducing atmospheric transparency. For ground-based optical-IR telescopes this effect is especially critical since it limits the quality of infrared observations.\\
To perform the analysis, we have considered the ERA5 variable \textit{total column water vapor}, $tcwv$. The variable is not a surface-level or vertically resolved variable, it's a single-level, vertically integrated value.\\
This parameter is the sum of water vapor, liquid water, cloud ice, rain and snow in a column extending from the surface of the Earth to the top of the atmosphere, thus it directly represents the total mass of water vapor in the column. The $tcwv$ and the PWV are closely related but have subtle differences, since the Precipitable Water Vapor represents the depth of water that would result if all the water vapor in the atmospheric column were condensed into liquid form. The PWV provides a more intuitive measure for meteorologists and astronomers, as it describes how much water would precipitate if all vapor condensed.\\
Authors in (\citenum{zhang2019}) have demonstrated an overestimation of the $tcwv$ with respect to direct measurements of PWV that could be carried out using radiometers or radiosonde at a given location. Results in (\citenum{zhang2019}) have indicated that the PWV derived from the ERA5 and ERA‐Interim show root‐mean‐square (RMS) errors of about 2.0 mm with respect to measurements with the Global Navigation Satellite System (GNSS), and about 3.0 mm with respect to measurements with radiosondes. The discrepancies vary between different epochs, as well as for different stations among the deployed network.\\
To provide a benchmark for comparison and improve the $tcwv$ calibration for the sites under analysis, data from the four sites in Saudi Arabia were supplemented with observations from Mount Armazones, the European Extremely Large Telescope (E-ELT) site on the northern Chilean Atacama coast, at an elevation of 3060 m a.s.l. Armazones is located in a desert environment with plateau-like topographic characteristics in its surroundings. For our case it represents is the closest comparison of a well documented site, having validated local measurements reported in the literature (\citenum{schock2009}; \citenum{otarola2010}), with a median PWV of 2.7 mm derived from a combination of ground-based and satellite observations (\citenum{otarola2010}).\\
\begin{figure}
\begin{center}
\begin{tabular}{c}
\includegraphics[height=9cm]{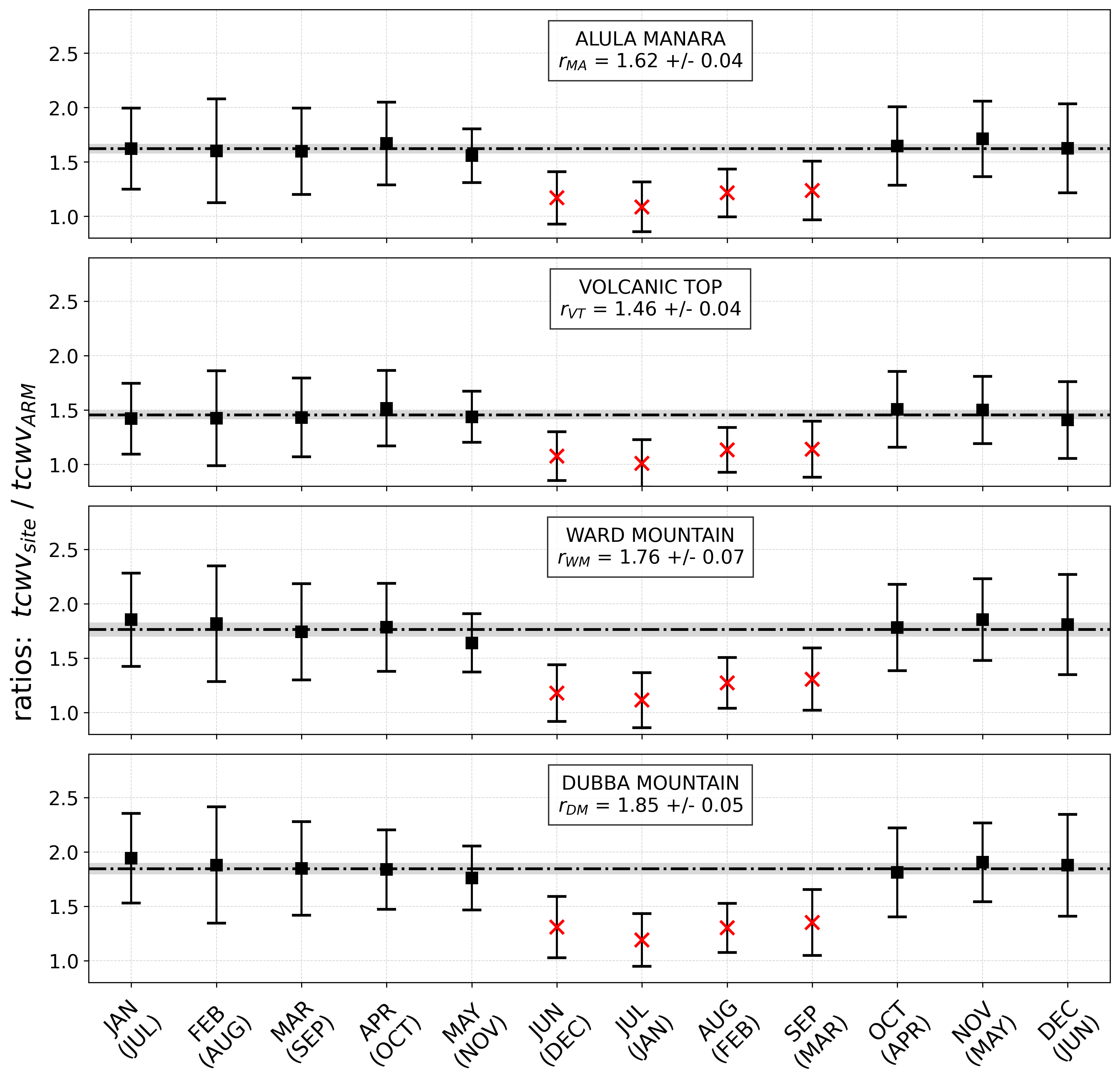}
\end{tabular}
\end{center}
\caption 
{\label{fig:tcwvratios} Ratios $r_{site} = tcwv_{site} / tcwv_{ARM}$. Armazones data are shifted by 6 months due to the inverted seasons with respect to the Saudi Arabia sites. The southern hemisphere months are reported in parenthesis below the northern hemisphere ones. The red \texttt{x} markers refer to ratios calculated during the Altiplanic Winter, and not accounted for the calculation of the averaged ratios. The dashed lines represents the averaged ratios, while the colored shade the relative standard deviations.} 
\end{figure} 
In Figure \ref{fig:tcwvratios} we show the comparison of the retrieved $tcwv$ from Armazones ($tcwv_{ARM}$) with those from our sites as the ratios $r_{site} = tcwv_{site} / tcwv_{ARM}$. It is important to notice that Armazones is located in the southern hemisphere, while the Saudi Arabia sites are located in the norther hemisphere, thus data from Armazones are shifted by 6 months due to the inverted seasons with respect to the Saudi Arabia sites.\\
Comparing Armazones with the other sites is not straightforward, since the climate pattern in the Chilean Atacama Andes is different from that in Western Saudi Arabia. During the southern hemisphere warm season, Armazones is affected by the so called "Altiplanic Winter", a specific climatic phenomenon that occurs in the high-altitude regions of the Andes, especially in countries like Bolivia, Argentina, Chile, and Peru. This period is marked by a cooling of the atmosphere over the Andean highlands, bringing rainy, cold weather to areas that are usually dry. During the Altiplanic Winter, moist air from the Amazon basin is pushed into the highlands of the Andes, which leads to heavy rain, cooler temperatures, and often snow at higher altitudes (\citenum{VicencioVeloso2024}). As a result, in those months the amount of water vapor in the atmosphere is higher with respect to drier months. Thus a comparison of Armazones with other sites must avoid using the Altiplanic Winter period. This is well spotted in Figure \ref{fig:tcwvratios}, where the red \texttt{x} markers refer to that period, and indeed the ratios are very different from the rest of the year, where, conversely, the ratios are very stable.\\
\begin{table}
\caption{Ratios $r_{site} = tcwv_{site} / tcwv_{ARM}$.} 
\label{tab:tcwvratios}
\begin{center}
\small
\begin{tabular}{|r|c|} 
\hline
  & $r_{site}$ \\
\hline\hline
\rule[-1ex]{0pt}{3.5ex} \textbf{AlUla Manara} & $1.62 \pm 0.04$\\  
\hline
\rule[-1ex]{0pt}{3.5ex} \textbf{Volcanic Top} & $1.46 \pm 0.04$\\  
\hline
\rule[-1ex]{0pt}{3.5ex} \textbf{Ward Mountain} & $1.76 \pm 0.07$\\  
\hline
\rule[-1ex]{0pt}{3.5ex} \textbf{Dubba Mountain} & $1.85 \pm 0.05$\\  
\hline
\end{tabular}
\end{center}
\end{table} 
The resulting ratios are reported in Table \ref{tab:tcwvratios}. AlUla Manara shows $tcwv$ values 62\% higher than Armazones, while Volcanic Top has drier conditions (46\% higher that Armazones). Surprisingly, the higher sites Ward Mountain and Dubba Mountain show higher content of $tcwv$ with respect to AlUla Manara. While AlUla Manara and Dubba Mountain lie on large plateaus, Ward Mountain and Dubba Mountain are prominent peaks (see Section \ref{sect:intro}). We believe that their relatively high $tcwv$ values may be overestimated due to the limited spatial resolution of the data (see Table \ref{tab:metrics}), which is insufficient to capture the expected decrease in $tcwv$ at the summit relative to the surrounding lower elevations.\\
Considering the proportional scaling correction to be applied to the Armazones $tcwv$ median value to match the 2.7 mm measured on site by (\citenum{otarola2010}), and supposing that the same RMS and scaling factor could affect the data for both the Saudi Arabia and Armazones sites, we could implement a common re-scaling for all sites. For Armazones this factor is $f = \text{med}(tcwv_{ARM})/2.7 = 5.3/2.7 = 2.56$, that means that the remote sensed $tcwv_{ARM}$ is measuring more than twice the local measured PWV. We intend to apply this correction factor to the $tcwv$ measured for the Saudi Arabia sites, and get a estimation of what we may expect from local measurements of the PWV. This method of calculating PWV from $f$ may not be accurate, and the $f$ factor will likely be very different at different sites, however, at this stage we only need to get a general direction while waiting real-world measurements from the radiometer instrument installed at the ASM (see Section \ref{sect:asm}).\\
\begin{table}
\caption{Calibrated median PWV prediction in [mm] for all sites on annual basis, and for the cold and warm season.} 
\label{tab:pwvmed}
\begin{center}     
\small
\begin{tabular}{|r|c|c|c|} 
\hline
\rule[-1ex]{0pt}{3.5ex}  & \textbf{Annual} & \textbf{Cold} & \textbf{Warm}  \\
\rule[-1ex]{0pt}{3.5ex}  &  & \textbf{Season} & \textbf{Season}  \\
\hline\hline
\rule[-1ex]{0pt}{3.5ex} \textbf{AlUla Manara} & 4.2 & 3.2 & 4.9  \\  
\hline
\rule[-1ex]{0pt}{3.5ex} \textbf{Volcanic Top} & 3.8 & 3.1 & 4.5  \\  
\hline
\rule[-1ex]{0pt}{3.5ex} \textbf{Ward Mountain} & 4.5 & 3.8 & 5.0  \\ 
\hline
\rule[-1ex]{0pt}{3.5ex} \textbf{Dubba Mountain} & 4.7 & 4.0 & 5.4  \\ 
\hline
\end{tabular}
\end{center}
\end{table} 
Table \ref{tab:pwvmed} reports the median calibrated PWV for the Saudi Arabia sites. Despite its much lower elevation of 1209 m.a.s.l. with respect to Armazones (3060 m.a.s.l.), AlUla Manara shows encouraging good conditions of 4.2 mm on an annual basis, and an estimated very good median of 3.2 mm during the cold season. In Figure \ref{fig:pwvseason} the median seasonal PWV averaged through the years from calibrated data is shown. Winter months have better conditions than summer months, and this is a common property between all sites.\\
\begin{figure}
\begin{center}
\begin{tabular}{c}
\includegraphics[width=1.0\textwidth]{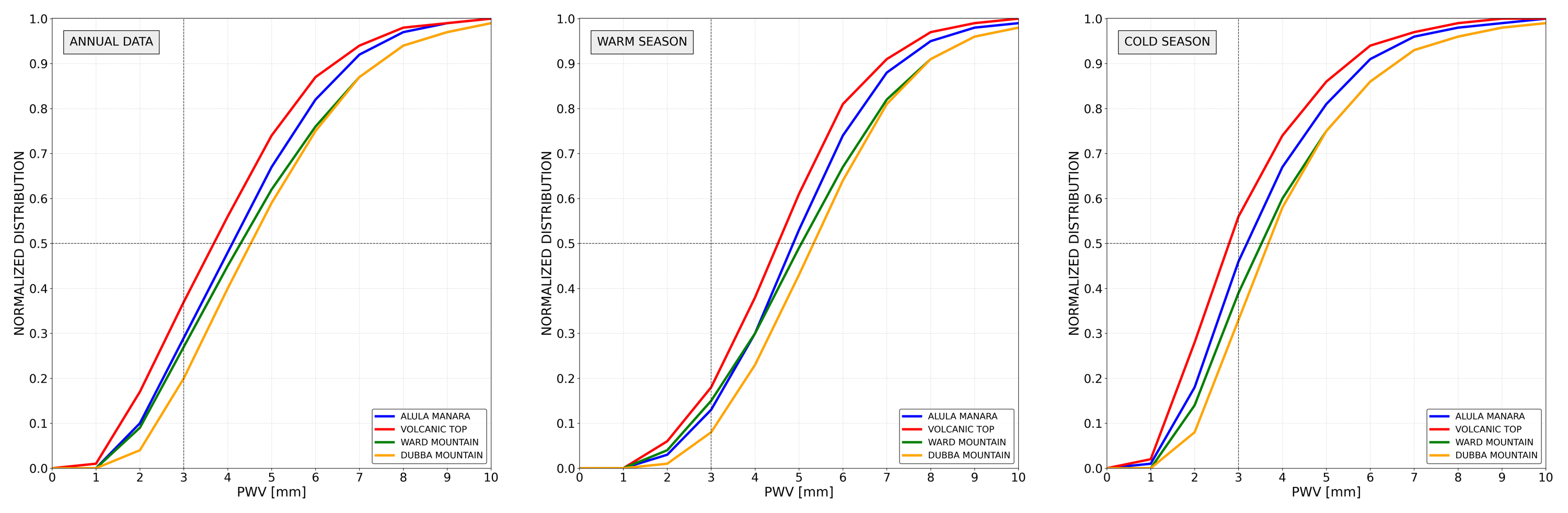}
\end{tabular}
\end{center}
\caption 
{\label{fig:pwvcumul} Calibrated PWV cumulative distributions in [mm] for AlUla Manara, Volcanic Top, Ward Mountain, and Dubba Mountain on annual basis (left), as well as for the warm season (center) and the cold season (right). The PWV = 3 mm line is marked to graphically highlight the estimated optimal amount of time available for outstanding infrared observations.} 
\end{figure} 
In Figure \ref{fig:pwvcumul} we show the calibrated PWV cumulative distributions on annual basis as well as for the cold season and the warm season. In the plots, the PWV = 3 mm line is marked to graphically highlight the estimated optimal amount of time available for outstanding infrared observations. In Table \ref{tab:pwvthermal} we report the fraction of nighttime hours having respectively PWV $<$ 3 mm or seeing in the visible $<$ 1.0 arcsec (0.51 arcsec in N-band). The statistics is shown on an annual basis, and for the warm and cold seasons. Results from Figure \ref{fig:pwvcumul} and Table \ref{tab:pwvthermal} indicate that on an annual basis, at AlUla Manara $\sim$30\% of nights are well suited for excellent infrared and mid-infrared observations. This percentage could rise to a remarkable $\sim$45\% of nights in the cold season, and a fair $\sim$13\% of nights in the warm season.\\
\begin{table}
\caption{Fraction of nighttime hours having respectively PWV $<$ 3 mm or seeing in the visible $<$ 1.0 arcsec (0.51 arcsec in N-band). Seeing ranges are retrieved from the Gm model in Section \ref{sect:seeing}. Numbers are reported on annual basis, and for the warm and cold seasons. The analysis is performed having in mind the performance of the sites hosting a 4m optical-IR telescope.}
\label{tab:pwvthermal}
\begin{center}
\small
\renewcommand{\arraystretch}{1.3}
\begin{tabular}{|>{\columncolor{gray!15}}c|r|c|c|c|} 
\hline
\cellcolor{white} & & \textbf{Annual} & \textbf{Warm} & \textbf{Cold}  \\
\cellcolor{white} & &  & \textbf{Season} & \textbf{Season}  \\
\hline
\noalign{\vskip 3pt}
\hline
 \textbf{AlUla} & PWV $<$ 3 mm & 29.5\% & 13.0\%  & 45.9\% \\
 \textbf{Manara} & seeing $<$ 1.0" & 20.1\% & 28.4\%  & 14.8\% \\ 
\hline
\noalign{\vskip 3pt}
\hline
 \textbf{Volcanic} & PWV $<$ 3 mm & 37.2\% & 18.4\% & 56.0\% \\
 \textbf{Top} & seeing $<$ 1.0" & 39.3\% & 57.2\%  & 23.3\% \\ 
\hline
\noalign{\vskip 3pt}
\hline
 \textbf{Ward} & PWV $<$ 3 mm & 26.9\% & 14.5\%  & 39.4\% \\
 \textbf{Mountain} & seeing $<$ 1.0" & 62.2\% & 84.4\%  & 40.5\% \\ 
\hline
\noalign{\vskip 3pt}
\hline
 \textbf{Dubba} & PWV $<$ 3 mm & 20.4\% & 10.1\% & 33.2\% \\
 \textbf{Mountain} & seeing $<$ 1.0" & 64.8\% & 87.4\% & 42.4\% \\
\hline
\end{tabular}
\renewcommand{\arraystretch}{1}
\end{center}
\end{table}
To put things into perspective, if we consider as an example the mid-infrared N-band having a central wavelength of 10 $\mu$m, at that wavelength a 4m optical-IR telescope has resolution $R_{N}$:
\begin{equation}
\label{eq:Nband}
R_N = 1.22 \frac{\lambda}{D} = 1.22 \frac{10\mu m}{4m} = 3.05 \times 10^{-6} \text{ radians} = 0.63 \text{ arcsec}
\end{equation}
while the diffraction limited FWHM of the telescope point-spread function (PSF) is expressed by
\begin{equation}
\label{eq:NbandFWHM}
FWHM = 0.98 \frac{\lambda}{D} = 0.98 \frac{10\mu m}{4m} = 2.45 \times 10^{-6} \text{ radians} = 0.51 \text{ arcsec}
\end{equation}
At AlUla Manara, a 1.5 arcsec nighttime median seeing in the visible wavelengths will correspond to about 0.8 arcsec in N-band. From the Gm modelling we retrieve a very good fraction of time of 28.4\% in the warm season, 14.8\% in cold season, and 20.1\% on annual basis, where the proposed 4m telescope will not be affected by the seeing (1.0 arcsec in the visible, $\sim$0.5 arcsec in N-band), producing diffraction limited images.\\
The calculated PWV content for AlUla Manara together with its estimated median seeing range are very promising, with values that, if confirmed, will set the possibility to perform high quality IR to mid-IR astronomical observations. The use of the upcoming  ASM Radiometer and the seeing monitor will confirm or discard these hypothesis (see Section \ref{sect:asm}).
\section{AlUla Manara Astronomical Site Monitor}\label{sect:asm}
The Astronomical Site Monitor for AlUla Manara has been designed following the classical scheme used for Very Large and Extremely Large Telescopes site characterization campaigns (\citenum{Lombardi2009}; \citenum{Lombardi2010}; \citenum{Lombardi2014}).\\
Recently, discussions have taken place concerning the possibility to expand the actual astronomical observatory project to a further solar telescope. The Kingdom of Saudi Arabia remains open to evaluating potential future expansions of the observatory, including the addition of a solar telescope. In this context, and to avoid any future gaps in data coverage, we have decided to implement continuous turbulence monitoring, both during the day and night. This ensures that, should solar observations or daytime dependent activities be pursued, the necessary site characterization data will already be available. Furthermore, seeing forecast models are also being actively studied, thus the availability of full 24-hour turbulence profiles will be beneficial for the model development. For the mentioned reasons, it has been decided to equip the ASM with a turbulence profiler capable of monitoring the turbulence during both daytime and nighttime.\\
\begin{figure}
    \centering
    \begin{tabular}{c}
        \includegraphics[width=0.85\textwidth]{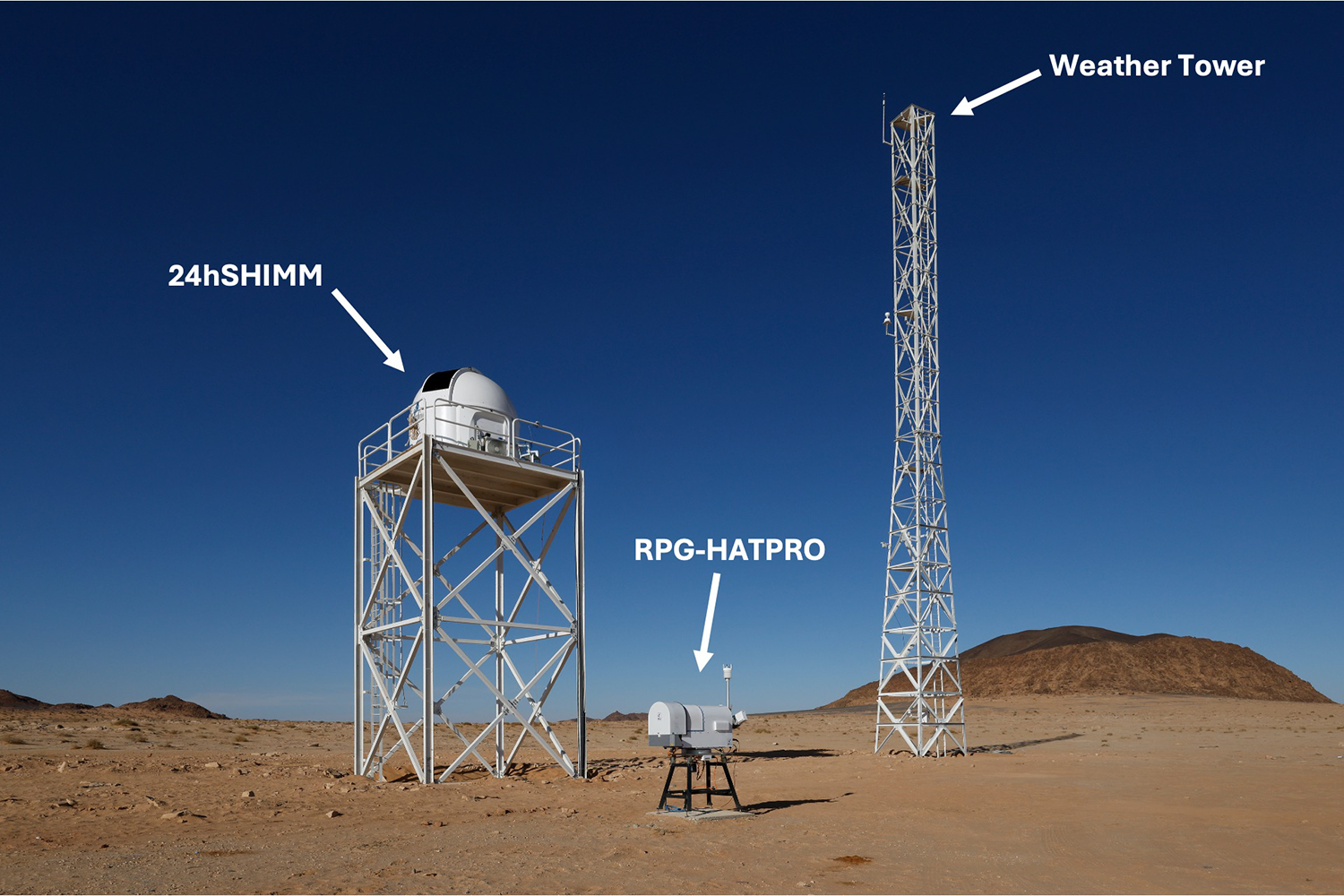}
     \end{tabular}
    \caption{\label{fig:asm} The AlUla Manara Astronomical Site Monitor on April 2025. The 24hSHIMM stands on the 8m tower on the left; the RPG-HATPRO Radiometer is placed on the ground, about 10 m from the 24hSHIMM tower; finally, the 30m Weather Tower is on the right, at about 30 m from the 24hSHIMM tower.}
\end{figure}
At the time of writing, the ASM civil works have been completed on site (Figure \ref{fig:asm}), and robotized operations began in late April 2025.
The AlUla Manara ASM is composed by the following instrumentation:
\begin{itemize}
    \item[$-$] \textbf{Atmospheric turbulence characterization}. A 24-hour Shack-Hartmann Image Motion Monitor (24hSHIMM; \citenum{Perera2016}; \citenum{Perera2023}; \citenum{Griffiths2023}; \citenum{Griffiths2024}) was chosen over a traditional Differential Image Motion Monitor (DIMM; \citenum{Lombardi2014}; \citenum{Sarazin1990}; \citenum{Tokovinin2002}) due to its capability to operate continuously, both during the day and at night (\citenum{Griffiths2023}). By utilizing a four-layer vertical turbulence profile and a wind speed profile obtained from meteorological forecasts, the 24hSHIMM can provide high-resolution estimates of key atmospheric parameters, including seeing, isoplanatic angle, coherence time, and Rytov variance. Given the anticipated extreme operational temperatures and significant day/night thermal gradients (see Section \ref{sect:t}), careful consideration has been given to its cooling and thermal management. To minimize the effects of surface layer turbulence, the instrument is being mounted on an Alt-Az mount on an 8-meter tower, thus its final elevation above ground will be of 10 m.
    \item[$-$] \textbf{Meteorological parameters}. Temperature, wind speed, wind direction, relative humidity, and atmospheric pressure will be continuously monitored for the entire lifespan of the future observatory. A 30-meter meteorological tower, equipped with Automated Weather Stations (AWS), has been designated and installed for permanent operations, with sensors positioned at the following heights above ground:
    \begin{itemize}
        \item 2 m, to monitor the ground conditions;
        \item 10 m, to monitor the conditions at the same level of the 24hSHIMM;
        \item 20 m, to monitor the conditions at an intermediate elevation;
        \item 30 m, to monitor the conditions close to the (expected) altitude of the 4m-class telescope dome.
    \end{itemize}
    \item[$-$] \textbf{Dust monitoring}. Given the desert arid conditions of the AlUla region, monitoring dust and aerosols in the atmosphere has been recognized as essential. To assess dust levels, particularly during the sporadic high wind events, an Air Particle Sensor is installed on-site and integrated into the AWS data system.
    \item[$-$] \textbf{Column of water vapor characterization}. An RPG-HATPRO radiometer is being deployed and integrated into the ASM to monitor the Precipitable Water Vapor. These measurements will enable the determination of the capabilities of the AlUla Manara Observatory for infrared astronomy.
\end{itemize}

\section{Conclusions}\label{sect:conclusions}
AlUla Manara, situated in the Medinah Province of northwest Saudi Arabia, represents a promising site for optical and infrared astronomy, aligning with the Kingdom’s Vision 2030 and its goals of integrating scientific research with sustainable tourism. Under the guidance of the Royal Commission for AlUla, a key initiative under this vision is the establishment of AlUla Manara as Saudi Arabia’s first major astronomical observatory, integrating astro-tourism with cutting-edge research through a 4m-class optical-IR telescope. Situated on a remote plateau 74 km North of the old Town of AlUla, AlUla Manara was recently designated as International Dark Sky Park, reinforcing its potential for astronomical research. This study presented a climatological and atmospheric assessment of the site using ECMWF ERA5 reanalysis data, with comparisons to nearby elevated locations outside the RCU jurisdiction, namely Volcanic Top, Ward Mountain, and Dubba Mountain.\\
Our findings indicate that AlUla Manara meets the criteria for hosting a 4m-class optical-IR telescope. The site exhibits a high fraction of clear nighttime skies exceeding 79\%, a median nighttime seeing of 1.5 arcsec, and a scaled precipitable water vapor of approximately 3.2 mm during the cold season, based on a correction applied to ERA5 total column water vapor values. Wind speeds are consistently low and do not pose significant challenges to telescope infrastructure.\\
Despite its lower elevation relative to alternative sites, which slightly limits turbulence performance, AlUla Manara remains a viable and well-balanced option. Special consideration must be given to thermal management, as the site experiences extreme temperatures, reaching up to 40\textdegree C during summer, both during the day and at night.\\
To enhance the accuracy of our assessment and guide future development, a dedicated Astronomical Site Monitor has been operational at AlUla Manara since April 2025. The continuous data it provides will be essential to refining turbulence models, validating remote sensing predictions, and supporting the telescope’s final design and operations plan. This integrated approach ensures that AlUla Manara can meet international standards for optical and infrared astronomical research while contributing to regional scientific capacity and outreach.

\section* {Acknowledgments}
The authors thanks the anonymous referees for contributing to the significant improvement of this paper. The authors also thank Rob Ivison at ESO for the useful comments on the early version of the manuscript, Angel Ot\'arola at ESO for the hints on the PWV measurements at Chilean sites, and James Osborn at University of Durham for facilitating contacts with LumiSpace in the context of the AlUla Manara ASM deployment. Omar Cuevas acknowledge support from Centro de Estudios Atmosf\'ericos y Cambio Clim\'atico (CEACC), Universidad de Valpara\'iso, Chile. Gianluca Lombardi acknowledges support from grant PID2022-137241NB-C41 funded by MICIU/AEI/10.13039/501100011033 and ERDF/EU.

%%%%% References %%%%%

\bibliography{report}   % bibliography data in report.bib
\bibliographystyle{spiejour}   % makes bibtex use spiejour.bst

\end{spacing}
\end{document}